\documentclass[twocolumn]{aastex63}
\usepackage{bm}
\usepackage{float}
\usepackage{footnote}
\usepackage{amsmath,amssymb}
\accepted{May 13, 2021}
\shorttitle{WASP-110b}
\shortauthors{Nikolov et al.}

\begin{document}


\title{{\emph{Ground-Based Transmission Spectroscopy with VLT\,\,FORS2:}}\\ Evidence for faculae and clouds in the optical spectrum of the warm Saturn WASP-110b}

\correspondingauthor{Nikolay Nikolov}
\email{nnikolov@stsci.edu}


\author[0000-0002-6500-3574]{Nikolay Nikolov}
\affiliation{Space Telescope Science Institute, 3700 San Martin Dr, Baltimore, MD 21218, USA}


\author[0000-0002-4195-5781]{Gracjan Maciejewski}
\affiliation{Institute of Astronomy, Faculty of Physics, Astronomy and Informatics, Nicolaus Copernicus University, ul. Grudziadzka 5, 87-100 Toru{\'n}, Poland}

\author[0000-0001-6839-4569]{Savvas Constantinou}
\affiliation{Institute of Astronomy, University of Cambridge, Madingley Road, Cambridge, CB3 0HA, UK}

\author[0000-0003-4328-3867]{Nikku Madhusudhan}
\affiliation{Institute of Astronomy, University of Cambridge, Madingley Road, Cambridge, CB3 0HA, UK}

\author[0000-0002-9843-4354]{Jonathan J. Fortney}
\affiliation{Department of Astronomy and Astrophysics, University of California, Santa Cruz, CA 95064, USA}

\author[0000-0002-3456-087X]{Barry Smalley}
\affiliation{Astrophysics Group, Keele University, Keele, UK}

\author[0000-0001-5365-4815]{Aarynn L. Carter}
\affiliation{Department of Astronomy and Astrophysics, University of California, Santa Cruz, Santa Cruz, CA 95064, USA}

\author[0000-0001-6391-9266]{Ernst J. W. de Mooij}
\affiliation{Astrophysics Research Centre, School of Mathematics and Physics, Queens University Belfast, Belfast BT7 1NN, UK}

\author{Benjamin Drummond}
\affiliation{Met Office, Fitzroy Road, Exeter, EX1 3PB, UK}

\author[0000-0002-9308-2353]{Neale P. Gibson}
\affiliation{School of Physics, Trinity College Dublin, Dublin 2, Ireland}

\author[0000-0003-4328-3867]{Christiane Helling}
\affiliation{Centre for Exoplanet Science, University of St Andrews, North Haugh, St Andrews, KY169SS, UK}
\affiliation{SUPA, School of Physics \& Astronomy, University of St Andrews, North Haugh, St Andrews, KY169SS, UK}
\affiliation{SRON Netherlands Institute for Space Research, Sorbonnelaan 2, 3584 CA Utrecht, NL}

\author[0000-0001-6707-4563]{Nathan Mayne}
\affiliation{Astrophysics Group, University of Exeter, Exeter, EX4 2QL, UK}

\author[0000-0001-5442-1300]{Thomas Mikal-Evans}
\affiliation{Kavli Institute for Astrophysics and Space Research, Massachusetts Institute of Technology, Cambridge, MA 02139, USA}

\author[0000-0001-6050-7645]{David K. Sing}
\affiliation{Department of Physics and Astronomy, Johns Hopkins University, Baltimore, MD 21218, USA}

\author{Jamie Wilson}
\affiliation{Astrophysics Research Centre, School of Mathematics and Physics, Queens University Belfast, Belfast BT7 1NN, UK}











\begin{abstract}
We present a ground-based optical transmission spectrum for the warm Saturn-mass exoplanet WASP-110b from two transit observations made with the FOcal Reducer and Spectrograph (FORS2) on the Very Large Telescope (VLT). The spectrum covers the wavelength range from $4000$ to $8333$\,\AA\, which is binned in 46 transit depths measured to an averaged precision of 220 parts per million (ppm) over an averaged 80\,\AA~bin for a~$\rm{{Vmag}}=12.8$ star. The measured transit depths are unaffected by a dilution from a close A-type field dwarf, which was fully resolved. The overall main characteristic of the transmission spectrum is an increasing radius with wavelength and a lack of the theoretically predicted pressure-broadened sodium and potassium absorption features for a cloud-free atmosphere. We analyze archival high-resolution optical spectroscopy and find evidence for low to moderate activity of the host star, which we take into account in the {\rm atmospheric} retrieval analysis. Using the AURA retrieval code, we find that the {\rm observed} transmission spectrum can be best explained by a combination of unocculted stellar faculae and a cloud deck.  Transmission spectra of cloud-free and hazy atmospheres are rejected at a high confidence. With a {\rm possible} cloud deck at its terminator, WASP-110b joins the increasing population of irradiated hot-Jupiter exoplanets with cloudy atmospheres observed in transmission.  \end{abstract}



\section{Introduction} \label{sec:intro}
Transmission spectroscopy is a powerful tool for constraining the composition and diversity of clear, cloudy and hazy atmospheres of irradiated gas-giant exoplanets. During a planetary transit, part of the starlight filters through the upper planetary atmosphere at the limb, causing the planetary effective radius to vary with wavelength (transmission spectrum), depending on the atmospheric composition \citep{seager00}. Transit observations from both space and the ground have started to reveal a prevalence of clouds and hazes in the atmospheres of these planets across the range of planetary mass, radius and temperature \citep{pont13, gibson13a, gibson13b, sing16}. Clouds and hazes have strong implications for all aspects of a planet's atmosphere and affect the observable spectra by effectively reducing the size of absorption features. Why some exoplanets have clear atmospheres while others have atmospheres dominated by hazes or clouds is currently not well understood. Enabling an exploration of the relationship between clouds/hazes and fundamental properties such as mass, radius, temperature, and composition requires a statistically-large sample of exoplanets, which has triggered multiple transmission spectroscopy efforts from space and the ground \citep{sing16, Sing19, Wyttenbach17, Palle17, Rackham2017, Huitson17, Kirk18}. 

In this paper, we present new results for the $4000-8333$\,\AA\,transmission spectrum of WASP110b from a large VLT transmission spectral survey with FORS2, comprising twenty exoplanets. This program rests on an earlier study demonstrating that FORS2 on the VLT is an ideal instrument for exoplanet characterization, capable of distinguishing between a clear, cloudy and hazy hot-Jupiter atmospheres \citep{nikolov16, gibson17, carter20}. Initial results from the present large VLT program have so far been presented in \cite{nikolov18b} for WASP-96b and \cite{wilson20} for WASP-103b, revealing a cloud-free atmosphere with sodium abundance constraint and a cloudy atmosphere, respectively. 



{\rm Detected by the} Wide Angle Search for Planets \citep{2006PASP..118.1407P}, WASP-110b is a sub-Jupiter transiting exoplanet (planetary mass \mbox{$M_{\rm{p}}=(0.510\pm0.064)~M_{\rm{J}}$}, where $M_{\rm{J}}$ is the mass of Jupiter, planetary radius \mbox{$R_{\rm{p}}=(1.238\pm0.056)~R_{\rm{J}}$}, where $R_{\rm{J}}$ is the radius of Jupiter, and equilibrium temperature \mbox{$T_{\rm{eq}}=1,134\pm33$~K)} orbiting a moderately bright \mbox{(Vmag$=12.762\pm0.092$)} G9 star (effective temperature \mbox{$T_{\rm{eff}}=5400\pm140$\,K,} surface gravity \mbox{$\log{g}=4.1\pm0.2$~(cgs)} and metallicity \mbox{$[{\rm{Fe/H}}]=-0.06\pm0.10$}) located at a distance of \mbox{$277\pm3$\,pc} in the south-eastern part of the constellation Sagittarius (\citealt{Anderson2014, Gaia2018}).

{\rm This paper} is organized as follows. Section\,\ref{sec:obs} and \ref{sec:calibs} detail our observations. We describe the data reduction and light curve analysis in Section\,\ref{sec:lcfits}. Results and discussion are presented in Section\,\ref{sec:discussion}.

\section{Observations}\label{sec:obs}
\subsection{VLT}\label{sec:vlt_obs}
\begin{figure*}[h]
\plotone{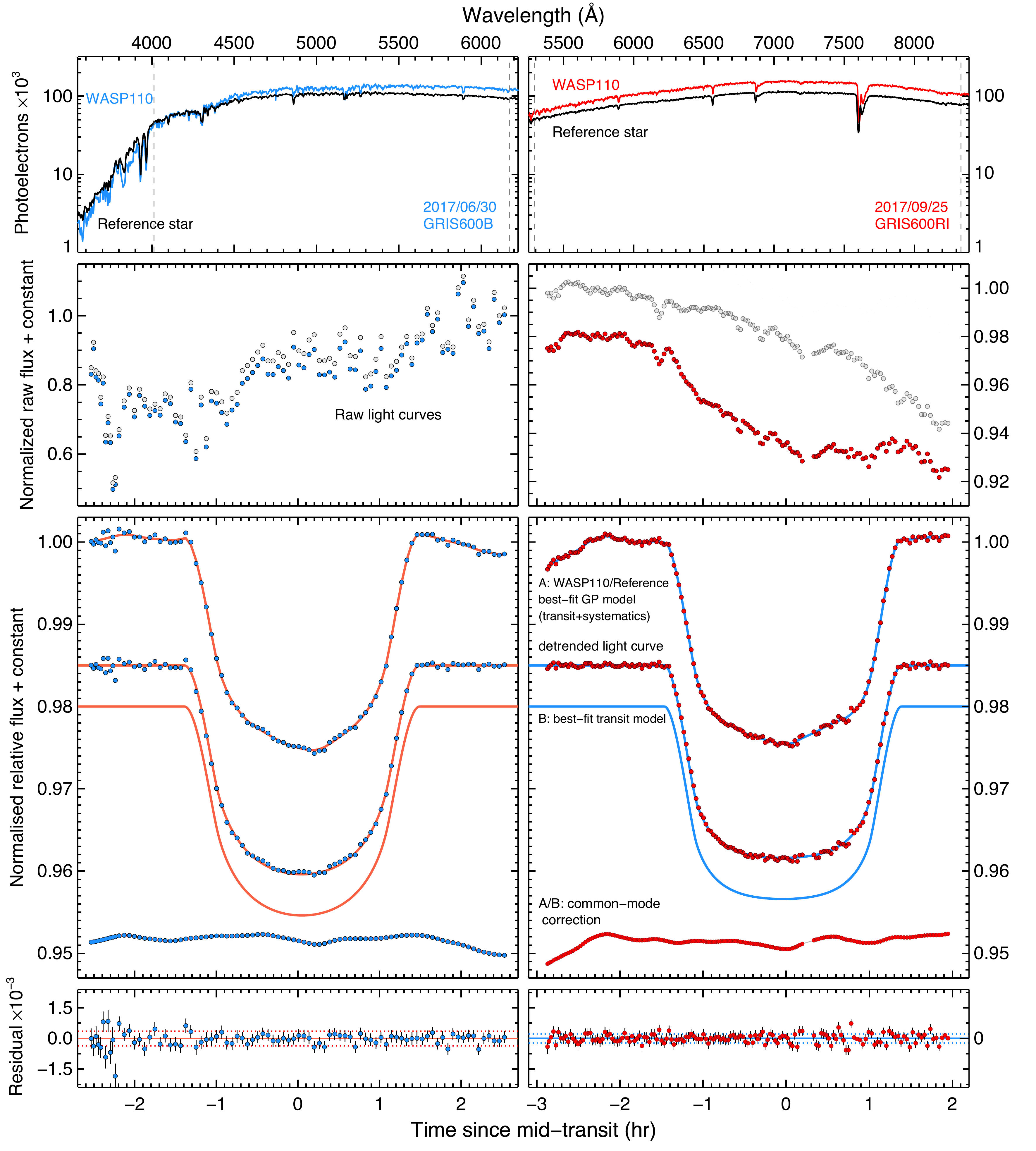}
\caption{VLT FORS2 stellar spectra and white-light curves compared to models. Left and right panels show the GRIS600B (blue) and GRIS600RI (red) datasets, respectively. The top row shows representative stellar spectra used for calibration with the dashed lines indicating the wavelength region used to produce the white-light curves. The second row shows normalized raw light curves for both target and reference star. The third row shows normalized relative target-to-reference raw flux along with the marginalized Gaussian process model (A), the detrended transit light curve and model (B), and the common-mode correction (A/B). The fourth row shows the best-fit light curve residuals and $1\sigma$ error bars, obtained by subtracting the marginalized transit and systematics models from the relative target-to-reference raw flux. The two light curve residuals show dispersions of 380 and 210 parts-per-million, respectively.   \label{fig:wlc}}
\end{figure*}

We observed two primary transits of WASP-110b on UT 2017 June 30th and September 25th with the FOcal Reducer and Spectrograph (FORS2, \citealt{Appenzeller1998}) attached to the Unit Telescope 1 (UT1, Antu) of the Very Large Telescope (VLT) at the European Southern Observatory (ESO) on Cerro Paranal in Chile {\rm{(Fig.\,\ref{fig:wlc})}} as part of Large Program 199.C-0467 (PI: Nikolov). We used similar observing setup and strategy to our earlier VLT studies (\citealt{nikolov16, gibson17, nikolov18b, carter20, wilson20}).




During the two transits, we collected data in multi-object spectroscopy mode with a mask consisting of two broad slits centered on the target and on a reference star of similar brightness. The reference star (known as 2MASS $20234536$-$4403456$) is a bright source in the FORS2 field of view Vmag$=12.999\pm0.013$ and is located at an angular separation of $2'.9$ eastward from the target and a distance of $604\pm14$\,pc from the Earth. The reference is of higher temperature \mbox{($T_{\rm{eff}} = 5912\pm126$\,K)} with magnitude differences (target minus reference) from the PPMXL and Gaia DR2 catalogues: $\Delta\,B=-0.36$, $\Delta\,G=-0.30$, $\Delta\,V=-0.24$, $\Delta\,R=-0.35$, $\Delta\,J=-0.46$, $\Delta\,H=-0.52$ and $\Delta\,K=-0.55$ (\citealt{roeser2010, Gaia2018}). We used broad slits spanning $22''$ along the dispersion and $\sim120''$ along the spatial (perpendicular) axis to minimize slit light losses due to seeing variations and guiding imperfections. We observed both transits with the same slit mask and the red detector (Massachusetts Institute of Technology - MIT), which is a mosaic of two chips. We positioned the instrument field of view such that each detector imaged one source. To improve the duty cycle, we made use of the fastest available read-out mode (200 kHz, $\sim30$s). During both nights, we ensured that the Longitudinal Atmospheric Dispersion Corrector (LADC) is in its neutral (park) position, i.e. inactive.

In addition to the bright reference star, a fainter star is located at $\sim5''$ eastward from WASP-110. The star (known as 6678937482610261632 in the Gaia DR2 catalog) is an A-type dwarf (T$_{\rm{eff}}=9617\pm128$\,K, M$\sim2.280M_{\odot}$, where M$_{\odot}$ is the mass of the Sun, R$\sim1.890R_{\odot}$, where R$_{\odot}$ is the radius of the Sun, $\log{g}\sim4.2$, luminosity L$\sim21$L$_{\rm{\odot}}$, where L$_{\rm{\odot}}$ is the luminosity of the Sun) of an apparent brightness of Vmag\,$=14.992\pm0.046$ and Gmag\,$=14.9710\pm0.0006$ at a distance of $4640\pm1237$\,pc and is not physically associated with the WASP-110 system (\citealt{Gaia2018}). The A-type dwarf is resolved from WASP-110 in the acquisition images and the spectroscopy time series of both nights.


During the first night we monitored the flux of WASP-110 and the comparison star under thin cirrus conditions. We used the dispersive element GRIS600B (hereafter blue), which covers the spectral range from $3600$ to $6200$\,\AA~at a resolving power $R = \lambda/\Delta \lambda \approx 600$. The field of view rose from an airmass $1.47$ to $1.06$ and at the end of the observations the field of view was at an airmass of $1.12$. The seeing gradually improved from $0.5''$ to $0.4''$, as measured from a Gaussian fit to the spatial profile of the stellar spectra. We collected a total of 87 exposures for $5$h with integration times of 80s and 200s, for the first 11 and remaining 76 exposures.

During the second night, we collected data under photometric conditions. We made use of the dispersive element GRIS600RI (hereafter red), which covers the range from $5400$ to $8200$\,\AA, in combination with the GG435 blocking filter to isolate the first order. The field of view rose from an airmass of $1.063$ to $1.060$ and the observations ended with the target at airmass of $1.853$. The seeing varied between $0.3''$ and $0.6''$. We monitored WASP-110 and the reference star for $4$h$49$m with $\sim8$\,min interruption at UT 3:33 due to a computer operational issue and collected a total of $154$ spectra with integration times between $80$ and $100$s.

\subsection{TESS}\label{sec:tess_obs}
WASP-110 was observed by the Transiting Exoplanet Survey Satellite \citep[TESS]{2014SPIE.9143E..20R} on Camera 1 CCD 2 during the Extended Mission between 2020 July 04 and 30 (Sector 27) with a 10-minute cadence in the full frame images (FFIs). A light curve was produced using tools available in the Lightkurve v1.9 package \citep{2018ascl.soft12013L} and applied to post-stamps $15 \times 15$ pixel ($5.25' \times 5.25'$) wide. They were extracted from FFIs with the TESSCut\footnote{https://mast.stsci.edu/tesscut/} online tool \citep{2019ascl.soft05007B} with WASP-110 centred in a frame. Fluxes were obtained employing a quadratic aperture which was 3-pixel wide without one corner pixel in order to avoid blending with a nearby bright star. The sky background level was determined with an algorithm based on standard-deviation thresholding. Trends caused by systematic effects or stellar variation were removed with the Savitzky-Golay filter with a window width of 12 hours. Prior to this step, in-transit and in-occultation data points were masked out using a trial ephemeris.

\begin{figure*}[ht]
\plotone{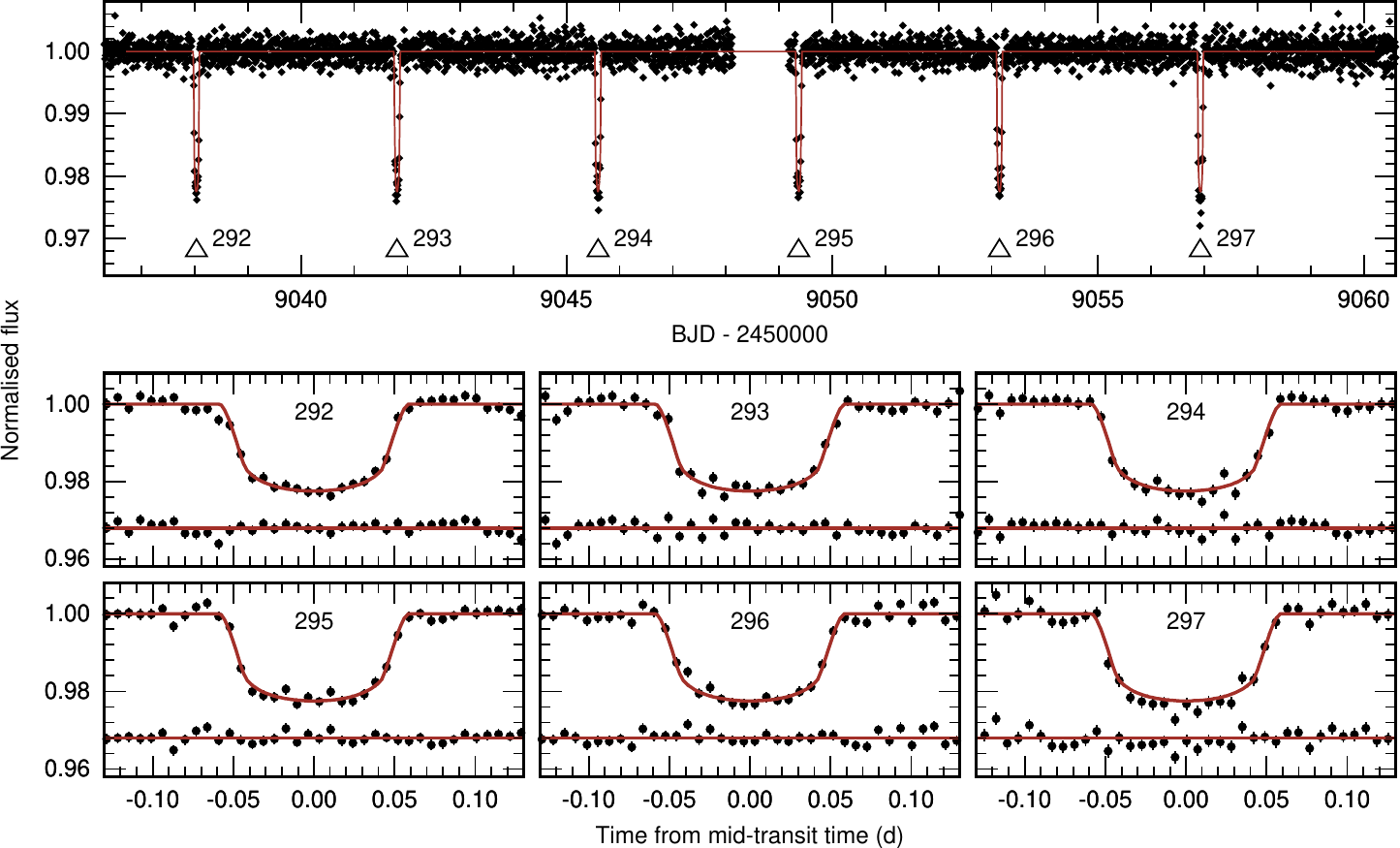}
\caption{Top panel: De-trended TESS light curve for WASP-110. The mid-transit times are indicated with open triangles and numbered following the ephemeris refined in this paper. The best-fitting transit model is drawn with a red line. Middle and lower panels: the individual transit light curves together with the best-fitting model. The residuals are plotted below each light curve and the vertical bars indicate the $1\sigma$ photometric uncertainties. \label{fig:tesslc}}
\end{figure*}

The final light curve is plotted in the upper panel in Fig~\ref{fig:tesslc}. Six consecutive transits of WASP-110~b were observed. For further analysis, the individual transit light curves were extracted with time margins equal to $\pm2.5$ times a transit duration. Their photometric noise rates \citep{2011AJ....142...84F} were found to be between 4.0 and 6.0 parts per thousand (ppth) of the normalised flux per minute of observation. Individual light curves are displayed in the middle and bottom panels in Fig~\ref{fig:tesslc}.

\section{Calibrations}\label{sec:calibs}
Our reduction and analysis {\rm{of the VLT FORS2 data}} commenced by subtracting a bias frame and by applying a flat field correction to the raw images. We computed a master bias and flat field by obtaining the median of 100 individual frames. We extracted one-dimensional spectra using the Image Reduction and Analysis Facility (IRAF)’s {\tt{APALL}} task by performing an unweighted summation. To trace the stars, we used a fit of a Legendre polynomial of two parameters. We removed the sky background by subtracting the median background from the stellar spectrum for each wavelength, computed from a box located away from the spectral trace. We found that aperture radii of 10 pixels and sky regions from 25 to 100 pixels (where the zero point is the middle of the spectrum special profile) minimize the dispersion of the out-of-transit flux of the band-integrated white light curves.

We performed a wavelength calibration of the extracted one-dimensional spectra using spectra of an emission lamp, obtained after each of the two transit observations with a mask identical to the science mask, but with slit widths of $1''$. We established a wavelength solution for each of the two stars with a second degree Legendre polynomial fit to the centres of a dozen lines, which we identified by performing a Gaussian fit. To account for displacements during the course of each observation and relative to the reference star, we placed the extracted spectra on a common Doppler-corrected rest frame through cross-correlation. All spectra were found to drift in the dispersion direction to no more than 2.5 pixels, with instrument gravity flexure and differential atmospheric dispersion being the most likely reasons.

Typical spectra of WASP-110 and the reference star are shown in Figure\,\ref{fig:wlc}. We achieved typical signal-to-noise ratios for the target and reference of 338 and 301 per pixel for the central wavelength of the blue grism and 354 and 306 for the red grism, respectively.

\section{Light curve analysis}\label{sec:lcfits}
Our light curve analysis of the VLT data closely follows the approach of \cite{nikolov18b}. For each transit, we produced wavelength-summed 'white' and spectroscopic light curves for WASP-110 and the reference star by summing the flux of each spectrum along the dispersion axis. We corrected the light curves for extinction caused by the Earth’s atmosphere by dividing the flux of the target by the flux of the reference star.

\begin{deluxetable}{lr}
\tablenum{1}
\tablecaption{System Parameters\label{tab:tab1}}
\tablewidth{0pt}
\tablehead{
\colhead{Parameter} & Value  
}
\startdata
Orbital period (day) & 3.7783977 (fixed)\\
Orbital eccentricity & 0 (fixed)\\
{\it{GRIS600B, $\lambda$}} (\AA) & 4013$-$6173 \\
$T_{\rm{mid}}$ (MJD) & $57934.23238^{+0.00026}_{-0.00028}$\\
$i$ ($^{\circ}$) & $88.00^{+0.45}_{-0.37}$\\
$a/R_{\ast}$ & $10.73^{+0.30}_{-0.35}$\\
$R_{p}/R_{\ast}$ & $0.1390^{+0.0063}_{-0.0064}$\\
$u_1$ & $0.52^{+0.22}_{-0.24}$\\
$u_2$ & $0.61^{+0.50}_{-0.47}$\\
$\ln{A}$ & $-13.0^{+3.3}_{-1.5}$\\
$\ln{\eta_{{\rm{time}}}}$ & $-0.9^{+2.7}_{-1.5}$\\
$c_0$ & $1.0000^{+0.0011}_{-0.0018}$\\
$c_1$ & $-0.00032^{+0.0010}_{-0.0007}$\\
{\it{GRIS600RI, $\lambda$}} (\AA) & 5293$-$8333 \\
$T_{\rm{mid}}$ (MJD) &  $58021.13912^{+0.00025}_{-0.00023}$\\
$i$ ($^{\circ}$) & $88.52^{+0.67}_{-0.46}$\\
$a/R_{\ast}$ & $11.19^{+0.21}_{-0.20}$\\
$R_{p}/R_{\ast}$ & $0.1389^{+0.0041}_{-0.0044}$\\
$u_1$ & $0.18^{+0.25}_{-0.22}$\\
$u_2$ & $0.74^{+0.40}_{-0.44}$\\
$\ln{A}$ & $-12.4^{+3.3}_{-1.3}$\\
$\ln{\eta_{{\rm{time}}}}$ & $-0.7^{+2.4}_{-1.2}$\\
$c_0$ & $1.0014^{+0.0013}_{-0.0027}$ \\
$c_1$ & $0.00097^{+0.0017}_{-0.00080}$\\
{\it{Weighted mean:}} & \\
$i$ ($^{\circ}$) & $88.16^{+0.37}_{-0.32}$\\
$a/R_{\ast}$ & $11.07^{+0.18}_{-0.17}$\\
{\it{GRIS600B}}  &  (fixed $i$, $a/R_{\ast}$, $T_{\rm{mid}}$) \\
$R_{p}/R_{\ast}$ & $0.1435^{+0.0047}_{-0.0036}$\\
$u_1$ & $0.66^{+0.14}_{-0.16}$\\
$u_2$ & $0.19^{+0.20}_{-0.19}$\\
{\it{GRIS600RI}}  & (fixed $i$, $a/R_{\ast}$, $T_{\rm{mid}}$) \\
$R_{p}/R_{\ast}$ & $0.1396^{+0.0033}_{-0.0037}$\\
$u_1$ & $0.12^{+0.20}_{-0.17}$\\
$u_2$ & $0.68^{+0.20}_{-0.24}$\\
\enddata
\end{deluxetable}

\subsection{White light curves}\label{sec:wlc_sec}
We divided  the raw flux of the target by the raw flux of the reference star to remove the variations of Earth’s atmospheric transparency, Figure\,\ref{fig:wlc}. We modeled the transit and systematic effects of the white-light curves by treating the data as a Gaussian process. \footnote{We made use of the publicly available {\tt{george}} Gaussian Process Python suite \citep{Foreman-Mackey15}.} and assuming quadratic limb darkening for the star (\citealt{gibson12a}). The transit parameters: mid-time $T_{\rm{mid}}$, orbital inclination $i$, normalized semi-major axis $a/R_{\ast}$ (where $R_{\ast}$ is the radius of the star), the planet-to-star radius ratio $R_{p}/R_{\ast}$ and the two limb-darkening coefficients $u_1$ and $u_2$ were allowed to vary in the fit to each of the two white-light curves, while the orbital period was held fixed to the previously determined value of \citet{Anderson2014}.

Under the Gaussian process assumption, the data likelihood is a multivariate normal distribution with a mean function $\mu$ describing the deterministic transit signal and a covariance matrix $K$ that accounts for stochastic correlations (i.e. poorly constrained systematics) in the light curves:

\begin{equation}
p(\bm{f} \vert \bm{\theta}, \gamma) = \mathcal{N}(\mu,K),
\end{equation}

where $p$ is the probability density function, $\bm{f}$ and $\bm{\theta}$ are vectors containing the flux measurements and mean function parameters, respectively; $\gamma$ is a function containing the covariance parameters and $\mathcal{N}$ is a multivariate normal distribution. The mean function $\mu$ is defined as:

\begin{equation}
\mu(\bm{t},\bm{\hat{t}};c_0,c_1,\bm{\theta})=[c_0+c_1\bm{\hat{t}}]\,T(\bm{\emph{t}},\bm{\theta}),
\end{equation}

where $\bm{t}$ is a vector of all central exposure time stamps in Julian Date, $\bm{\hat{t}}$ is a vector containing all standardized times, that is, with subtracted mean exposure time and divided by the standard deviation, $c_0$ and $c_1$ describe a linear baseline trend, $T(\bm{\theta})$ is an analytical expression describing the transit and $\bm{\theta}=(i, a/R_{\ast}, T_{\rm{mid}}, R_{p}/R_{\ast}, u_1, u_2)$. We made use of the analytical formulae of \citet{mandel02, kreidberg2015}.

We computed the {\rm{theoretical}} values of the coefficients of the quadratic limb darkening law using a three-dimensional stellar atmosphere model grid \citep[][]{magic2015}, factored by the throughputs of the blue and red grisms. In these calculations, we adopted the closest match to the effective temperature, surface gravity and metallicity of the exoplanet host star found in \citet{Anderson2014}. The choice of a quadratic versus a more complex law (such as a four-parameter nonlinear law) was motivated by the study of \citet{espinoza2016}, in which the two-parameter law has been shown to introduce negligible bias on the measured properties of transiting systems similar to WASP-110. The quadratic law also requires a much shorter computational time to compute model transit light curve. {\rm{The theoretical limb darkening coefficients were employed as priors in our light curve fits.}}

Similar to our earlier VLT studies, we defined the covariance matrix as $K = \sigma_i^2\delta_{ij} + k_{ij}$, where $\sigma_i$ contains the photon noise uncertainties, $\delta_{ij}$ is the Kronecker delta function and $k_{ij}$ is a covariance function. The white noise term $\sigma_w$ was assumed to have the same value for all data points and was allowed to freely vary. We chose to use the Mat\'{e}rn $\nu = 3/2$ kernel with time $t$ for the covariance function. Our kernel choice is motivated by the study of \cite{gibson13b}, where the Mat\'{e}rn $\nu = 3/2$ kernel is empirically motivated using simulated data, and is the first to use this kernel for light curve analysis. We also experimented by adding additional terms, including the spectral dispersion and cross-dispersion drifts $x$ and $y$ as input variables, and the full-width at half-maximum (FWHM) measured from the cross-dispersion profiles of the two-dimensional spectra and the speed of the rotation angle $z$. As with the linear time term, we also standardized the input parameters before the light curve fitting. We chose to use the time for both observations instead of combinations of other terms, because a GP of time resulted in well-behaved residuals. The covariance function was defined as:

\begin{equation}
k_{ij} = A^2(1+\sqrt{3}D_{ij})\exp{-\sqrt{3}D_{ij}},
\end{equation}

where $A$ is the characteristic correlation amplitude and

\begin{equation}
D_{ij} = \sqrt{\frac{(\bm{\hat{t}}_i-\bm{\hat{t}}_j)^2}{\tau_t^2}},
\end{equation}

where $\tau_{t}$ is the correlation length scale and the hatted variables are standardized. The parameters $\bm{X} = (c_0, c_1, T_{\rm{mid}}, i, a/R_{\ast}, R_{p}/R_{\ast}, u_1, u_2)$ and $\bm{Y} = (A, \tau_t)$ were allowed to vary and fixed the orbital period $P$ to its literature value. Uniform priors were adopted for $\bm{X}$ and log-uniform priors for $\bm{Y}$.

We marginalized the posterior distribution $p(\theta,  \gamma | f ) \varpropto p(f | \theta,  \gamma)p(\theta, \gamma)$ using the Markov chain Monte Carlo software package emcee \citet{Foreman13}. We identified the maximum likelihood solution using the Levenberg–Marquardt least-squares algorithm \citep[][]{markwardt09} and initialized three groups of $150$ walkers close to that maximum. The first two groups were run for $350$ samples and the third one for $4,500$ samples. To ensure faster convergence, we re-sampled the positions of the walkers in a narrow space around the position of the best walker from the first run before running for the second group. This helps prevent some of the walkers starting in a low-likelihood area of parameter space, which can require more computational time to converge. Figure\,\ref{fig:wlc} and Table\,\ref{tab:tab1} show transit models for each of the two observations computed using the marginalized posterior distributions and the fitted parameters, respectively. We find residual dispersion of 378 and 223 parts per million for the blue and red light curves, respectively. 

We computed the weighted mean values of $i$ and $a/R_{\ast}$ and repeated the fits allowing only $R_{p}/R_{\ast}$, $u_1$ and $u_2$ to vary. We held fixed $i$ and $a/R_{\ast}$ to the weighted mean values and the two central times, $T_{\rm{mid}}$ were held fixed to the values determined from the first fit. We report our results in Table\,\ref{tab:tab1}.

\begin{figure}[!ht]
\begin{centering}
\includegraphics[width=0.45\textwidth]{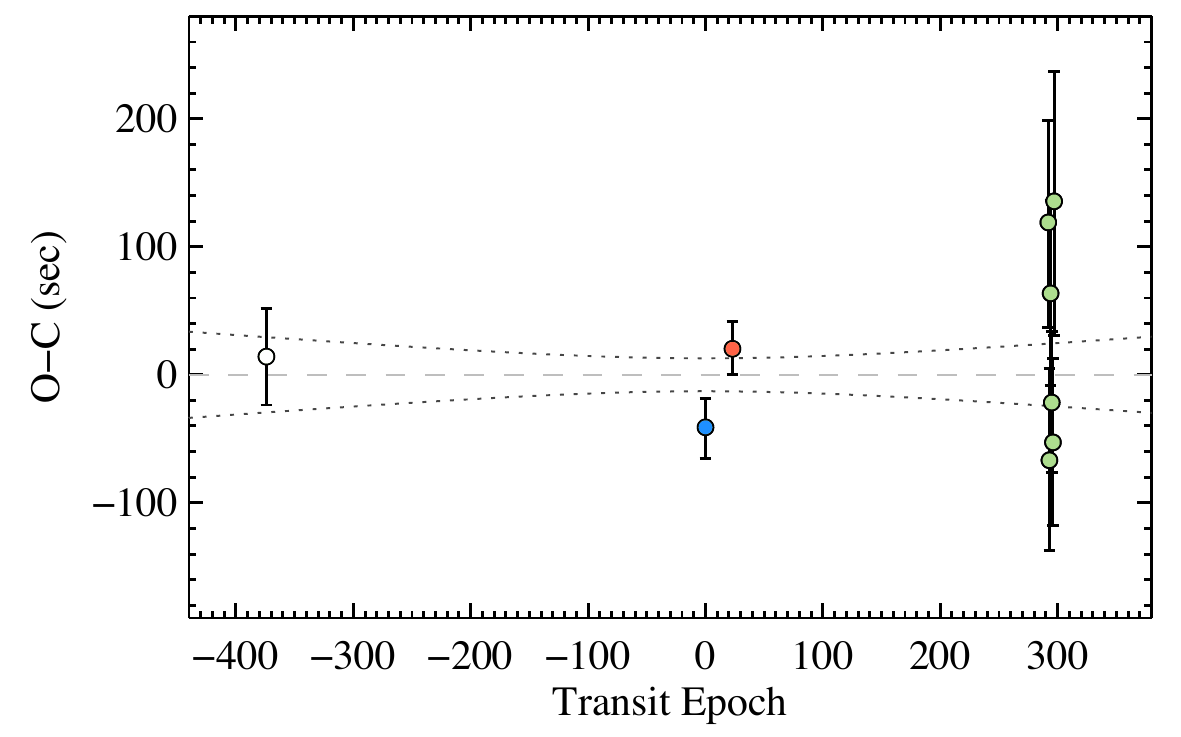}
\caption{Observed minus computed (O-C) transit times based on the best-fitting orbital period and central transit time derived from our analysis of the Euler (white dot), VLT (blue and red dots) and TESS (green dots) light curves. The error bars and dotted lines indicate the $1\sigma$ uncertainties of the transit times and derived ephemeris, respectively. \label{fig:oc}}
\end{centering}
\end{figure}

\begin{deluxetable}{ccc}
\tablenum{2}
\tablecaption{Central transit times\label{tab:omc}}
\tablewidth{0pt}
\tablehead{
\colhead{Epoch} & Central time, ${\rm{BJD}}_{{\rm{TDB}}}$ & O$-$C, day
}
\startdata
 -374 & $242456521.61682\pm{ 0.00044}$ &   0.000164\\
    0 & $242457934.73824\pm{ 0.00028}$ &  -0.000475\\
   23 & $242458021.64217\pm{ 0.00023}$ &   0.000235\\
  292 & $242459038.03324\pm{ 0.00095}$ &   0.001378\\
  293 & $242459041.80949\pm{ 0.00081}$ &  -0.000773\\
  294 & $242459045.58940\pm{ 0.00084}$ &   0.000736\\
  295 & $242459049.36682\pm{ 0.00063}$ &  -0.000251\\
  296 & $242459053.14486\pm{ 0.00075}$ &  -0.000611\\
  297 & $242459056.9254\pm{ 0.0012}$ &   0.001569\\
\enddata
\end{deluxetable}

\subsection{TESS light curves and transit ephemeris}\label{sec:tess_lcs}

The TESS light curves were used to refine the transit ephemeris for WASP-110b. In addition, we used a follow-up photometric time series from the discovery paper \citep{2014arXiv1410.3449A}. This early epoch observation was acquired on 2013 August 16 with the 1.2-m Euler-Swiss telescope equipped in the EulerCam photometer. In total, 121
measurements were collected in a Cousins I filter with a median cadence of 137 s. The photometric noise rate was found to be 1.15 ppth per minute of observation. The original timestamps in $\rm{HJD_{UTC}}$ were converted into $\rm{BJD_{TDB}}$ using an online applet\footnote{http://astroutils.astronomy.ohio-state.edu/time/hjd2bjd.html} \citep{2010PASP..122..935E}. For timing purposes, we do not correct the TESS transit light curves for the third-light contamination. Thus, we used fluxes prior the dilution correction applied in \citep{2014arXiv1410.3449A}.

The mid-transit times were determined by fitting a trial transit model with the Transit Analysis Package (TAP, \citet{2012AdAst2012E..30G}). The TESS and EulerCam light curves were modelled simultaneously with the orbital inclination, the semi-major axis scaled in star radii, and the ratio of planet to star radii linked together. The limb darkening (LD) effect was approximated with a quadratic law, the coefficients of which were bi-linearly extrapolated from tables of \citet{2011A&A...529A..75C}. The coefficients for the TESS passband were calculated as averages of $R$, $I$, and Sloan Digital Sky Survey $z'$ values. Their values were allowed to vary under the Gaussian penalties with a conservative width of 0.1. Possible trends in the time domain were accounted for using second order polynomials for each transit light curve. The best-fitting values and their uncertainties were calculated as the medians and 15.9 and 84.1 percentiles of the marginalised posteriori probability distributions generated from 10 MCMC chains, each $10^6$ steps long with 10\% burn-in phase. The results are given in Table\,\ref{tab:omc}.

We converted to BJDs the central transit times from our VLT observations and combined these with the BJD TDB central times from our Euler and TESS light curve analysis and computed an updated transit ephemeris. We fitted a linear function of the orbital period ($P$) and transit epoch ($E$): $T_{\rm{C}}(E)=T_{0}+EP$. We find a period of $P = 3.77840121\pm0.00000082$\,(d) and a mid-transit time of $T_{\rm{C}} = 2457934.73871\pm 0.00015$\,(d). The updated ephemeris is in agreement with the ephemeris reported by \cite{Anderson2014} and does not indicate the presence of transit timing variations (TTVs) Figure\,\ref{fig:oc}.

\subsection{Spectroscopic light curve analyses}
To produce spectroscopic light curves, we summed the flux of the target and reference star in bands with variable widths from 80 to 253\,\AA, Table\,\ref{tab:trspectab}. The blue and red grisms overlap in the wavelength range from $5,300$ to $6,200$\,\AA\,\,and each covers the sodium D lines at $5,890$ and $5,896$\AA. We choose common bins within the overlapping region to allow a direct comparison when combining the blue and red transmission spectra. We enlarged the first three bins in the blue grism and merged two pairs of bins, covering the O$_2$ A and B telluric bands from $7,594$ to $7,621\,$\AA\,\,and from $6,867$ to $6,884$\,\AA, respectively. In this way, we increased the signal-to-noise ratio of the broader bands and produced a total of {\rm{46}} light curves.

We established wavelength-independent, i.e. common mode systematic correction factors for each night, similar to our previous studies with FORS2 (\citealt{nikolov16, gibson17, nikolov18b, carter20, wilson20}). To obtain the correction factors, we divided the white-light curve from each night by the transit model computed with the weighted-mean system parameters (Section\,\ref{sec:wlc_sec}). The common mode factors for each night are shown in Figure\,\ref{fig:wlc}. 

We modeled the spectroscopic light curves with a function accounting for the transit and systematics simultaneously. Before fitting, we divided each spectroscopic light curve by the wavelength-independent systematic correction factors for that night. In the fits, we allowed only the relative planet radius, $R_{p}/R_{\ast}$ and the linear limb-darkening coefficient $u_1$ to vary. {\rm{We obtained theoretical values for $u_{1}$ and $u_{2}$ following the same approach as for the white light curves.}} We also fitted for both limb-darkening coefficients and found that the uncertainty of $u_2$ is large and consistent with the theoretical prediction. We interpret this as an indication for insufficient constraining power of the data for the {\rm{quadratic}} coefficient. Given the transmission spectra did not substantially change we chose to fix $u_2$ and to fit only for $u_1$. 

We accounted for the systematics using a low-order polynomial (up to a second degree with no cross terms) of dispersion and cross-dispersion drift, air mass, FWHM and the rate of change of the rotator angle. We produced all possible combinations of detrending variables and performed separate fits including each combination within the systematics function. This approach has been prefered as opposed to GP regression, as the CM-corrected VLT spectroscopic light curves exhibit a lower level of systematic effects. Our choice also rests on results from a VLT comparative follow-up of WASP-39b with atmospheric features detected with the {\it{Hubble Space Telescope}} \citep[][]{nikolov16}. We computed the Akaike Information Criterion for each fit and estimated the statistical weight of the model depending on the number of degrees of freedom, (\citealt{akaike74}). We chose to rely on the Akaike Information Criterion instead of other information criteria, for example, the Bayesian Information Criterion (\citealt{Schwarz78}), because the Akaike Information Criterion selects more complex models, resulting in more conservative error estimates. We marginalized the resulting relative radii and linear limb-darkening coefficient following \citealt{gibson14}. We found {\rm{that for most light curves}} systematics models parameterized with linear air mass, dispersion drift and FWHM terms {\rm{resulted}} in the highest statistical weight.


Before fitting each light curve, we set the spectrophotometric uncertainties of each band to values that include photon and readout noise. To determine the best-fit models, we used a Levenberg–Marquardt least-squares algorithm and rescaled the uncertainties of the fitted parameters using the dispersion of the residuals, \citep[][]{markwardt09}. Residual outliers larger than $3\sigma$, typically 2-3 per light curve, were excluded from the analysis. Correlated residual red noise was accounted following the methodology of \cite{pont06} by modeling the binned variance with the relation $\sigma^2 = \sigma_w^2/N + \sigma_r^2$ relation, where $\sigma_w$ is the uncorrelated white noise component, $N$ is the number of measurements in the bin and $\sigma_r$ is the red noise component. {\rm{We found white noise dispersion in the range from about 450 to 900 parts per million. For the red noise, we found a dispersion in the range from about 40 to 70 parts per million.}}


\subsection{Stellar Activity and Variability Monitoring}\label{sec:activity}
Activity and rotation are known to complicate the interpretation of transmission spectra and in particular when combining multi-instrument, multi-epoch data sets \citep[][]{huitson13, McCullough14, Pinhas2018}. To asses the level of stellar activity and photometric variability, we inspected the cores of the Ca\,{\sc{II}}\,H\&K lines and photometry time series from archival data. We analyzed five MPG/ESO 2.2-metre telescope high-resolution spectra obtained with the FEROS spectrograph (ESO program: 099.A-9010, PI P. Sarkis) at a resolution of $R=\lambda/\delta\lambda=48,000$, Table\,\ref{tab:feros}. The spectra have been obtained using a fiber with a diameter of 2\,arcsec and reduced with ESO's FEROS pipeline to phase 3 products, i.e. extracted one-dimensional spectra with wavelength solution. {\rm{A comparison between the tabulated and observed wavelengths of the Ca\,{\sc{II}}\,H\&K line cores indicates an offset (Figure\,\ref{fig:feros})}}. This can be attributed to the low signal-to-noise of the individual spectra, which reach $\sim5$ in the blue wavelengths, but can also be an indication for the presence of emission in the core of the lines. The activity index, $S_{\rm HK}$, was estimated from the spectrum using the method described by \cite{1978PASP...90..267V} and transformed to the $\log R'_{\rm HK}$ system using the relations from \cite{1984ApJ...279..763N}.  We measure $\log{R'_{\rm HK}}=-4.9^{+0.09}_{-0.24}$, which is consistent with low to moderate activity of the host star.

\begin{figure}[h]
\centering
\includegraphics[width=0.47\textwidth]{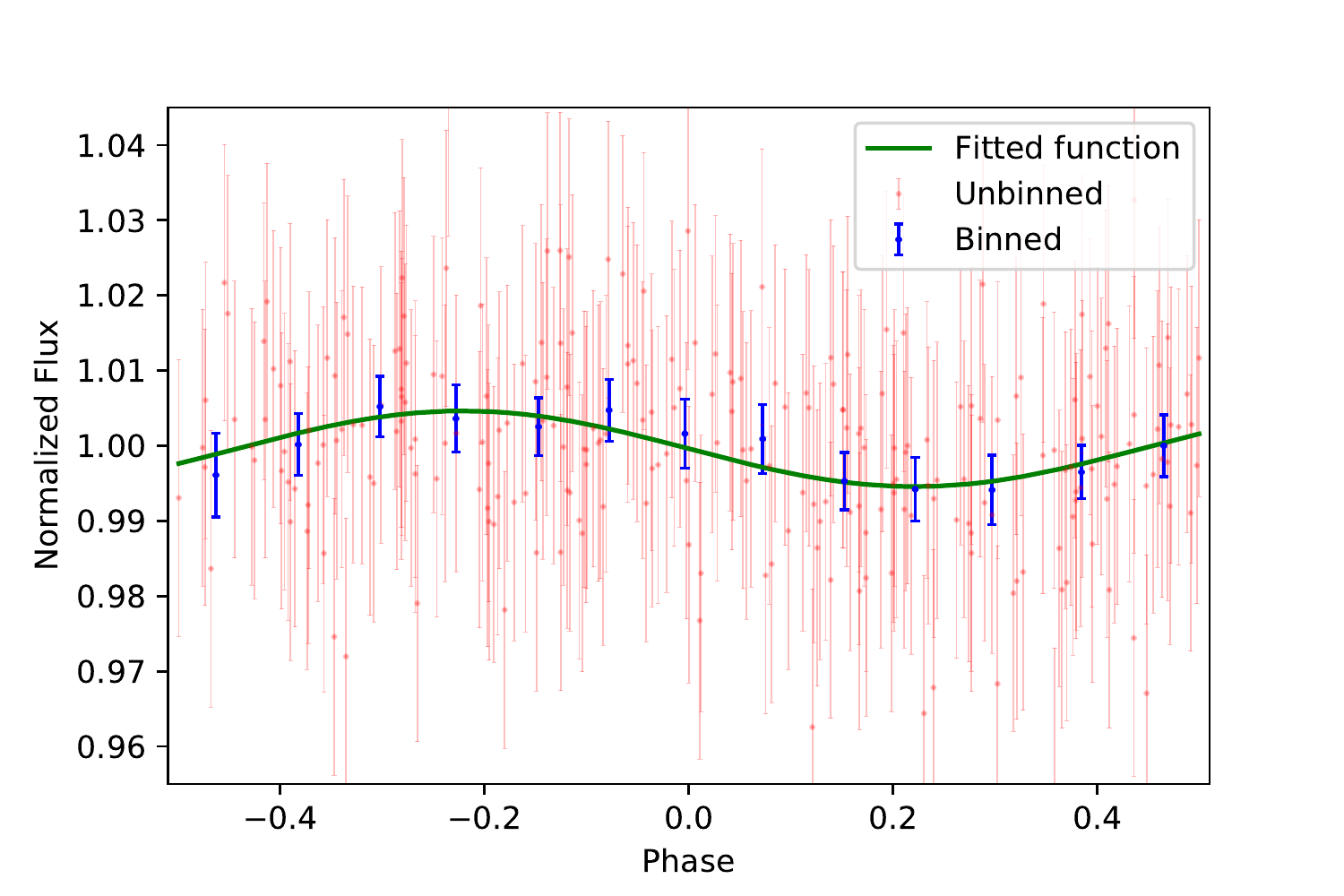}
\caption{{\rm{ASAS-SN photometry phase-folded with the highest significance period found from our periodogram analysis. Plotted are the unbinned photometry (red), binned (blue) and the best-fit sine function (green). The vertical bars represent $1\sigma$ uncertainties.}} \label{fig:asassn_folded}}
\end{figure}

\begin{figure}[h]
\centering
\includegraphics[width=0.47\textwidth]{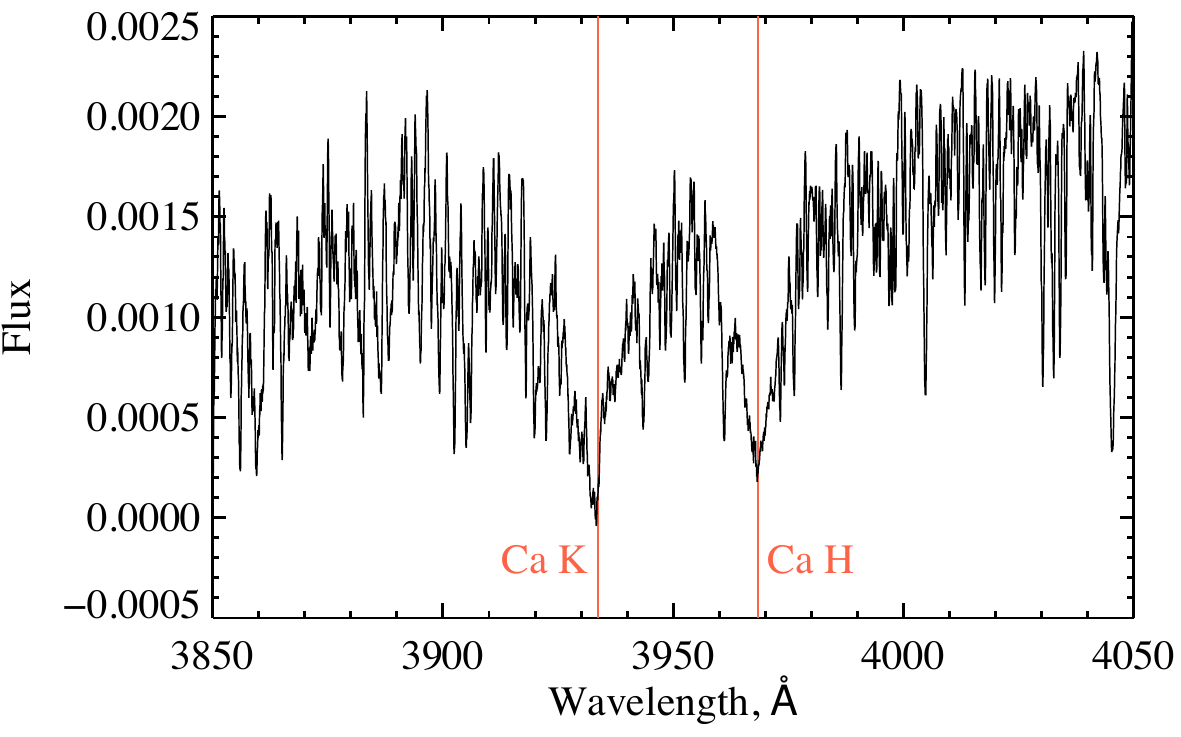}
\caption{Combined FEROS spectra around the Ca\,{\sc{II}}\,H\&K lines for WASP-110. The laboratory central wavelength of each line are indicated with the red continuous lines.  \label{fig:feros}}
\end{figure}

We analyzed an archival light curve consisting of 250 photometric data points from the Ohio State University’s All-sky Automated Survey for Supernovae (ASAS-SN) Photometry Database, \citet{Shappee2014, Jayasinghe2019}. The data has been obtained during the period from May to December 2014 and from March to December in the years from 2015 to 2018, respectively (Figure~\ref{fig:varLC1}). {\rm{We performed Lomb-Scargle periodogram analysis with trial frequencies ranging between 0.005 and $1\,\rm{d^{-1}}$, corresponding to a period range between 1 and 200\,d. We find a rotational modulation with a period of $13.74$\,d with a false alarm probability (FAP) of $\sim0.25$. We phase-folded the photometry and fitted a sinusoid to measure the amplitude of the flux variation. We found an amplitude of $0.5\pm0.1\%$ and residual scatter of 12.8 parts-per-thousand (ppt), Figure\,\ref{fig:asassn_folded}. Our result agrees with the 4\,mmag ($0.4\%$) upper limit for flux variability of WASP-110 reported by \cite{Anderson2014}. \cite{Anderson2014} also found that WASP-110 is an evolved $8.6\pm3.5$\,Gyr G9 star, which implies a gyrochronologic rotation comparable or exceeding the equatorial rotation period of the Sun i.e., $\gtrsim25$\,days. \cite{Anderson2014} report $v\sin(i_{\ast})=0.2\pm0.6$\,${\rm{kms^{-1}}}$, which translates to a rotation period $P_{\ast}(i_{\ast}=90^{\circ})\gtrsim34$\,days. While this is a discrepancy with our finding, a detailed comparison is hampered by the unconstrained stellar axial tilt, which can significantly decrease the actual rotation period. It is worth mentioning that empirical relations of the rotation periods to the chromospheric activity level $\log{R'_{\rm HK}}$ predict $\sim22$\,days for WASP-110 \citep[their Figure 13]{2015MNRAS.452.2745S}. Given the spread of $\sim10$\,days in the empirical relationship for the rotation period for G-type stars with $-0.1<\left[{\rm{Fe/H}}\right]<0.1$, the predicted estimate is consistent with our measured period. This is also the case for $v\sin(i_{\ast}\sim4^{\circ})$ of \cite{Anderson2014}. }}

\begin{deluxetable}{ccccc}
\tablenum{3}
\tablecaption{FEROS observing log\label{tab:feros}}
\tablewidth{0pt}
\tablehead{
\colhead{Date} & Exposure & Airmass & Seeing &  $\rm{S}/\rm{N}$ in \\
\colhead{2017 (UT)}               & time (s)          &              & (arcsec) 	& $0.03$\AA\,\,bin
}
\startdata
 May 14 & 3,000 & 1.209&   1.72 & 24.2 \\
 May 17 & 2,400 & 1.178&   $<0.91$ & 24.8 \\
 June 02 & 1,200 & 1.042 & 1.15 & 27.7 \\
 July 21 & 1,200 & 1.035 & 0.92 & 25.4 \\
 July 21 & 1,200 & 1.035 & 0.91 & 25.2 \\
\enddata
\end{deluxetable}

\subsection{Results} 
\begin{figure}
\centering
\plotone{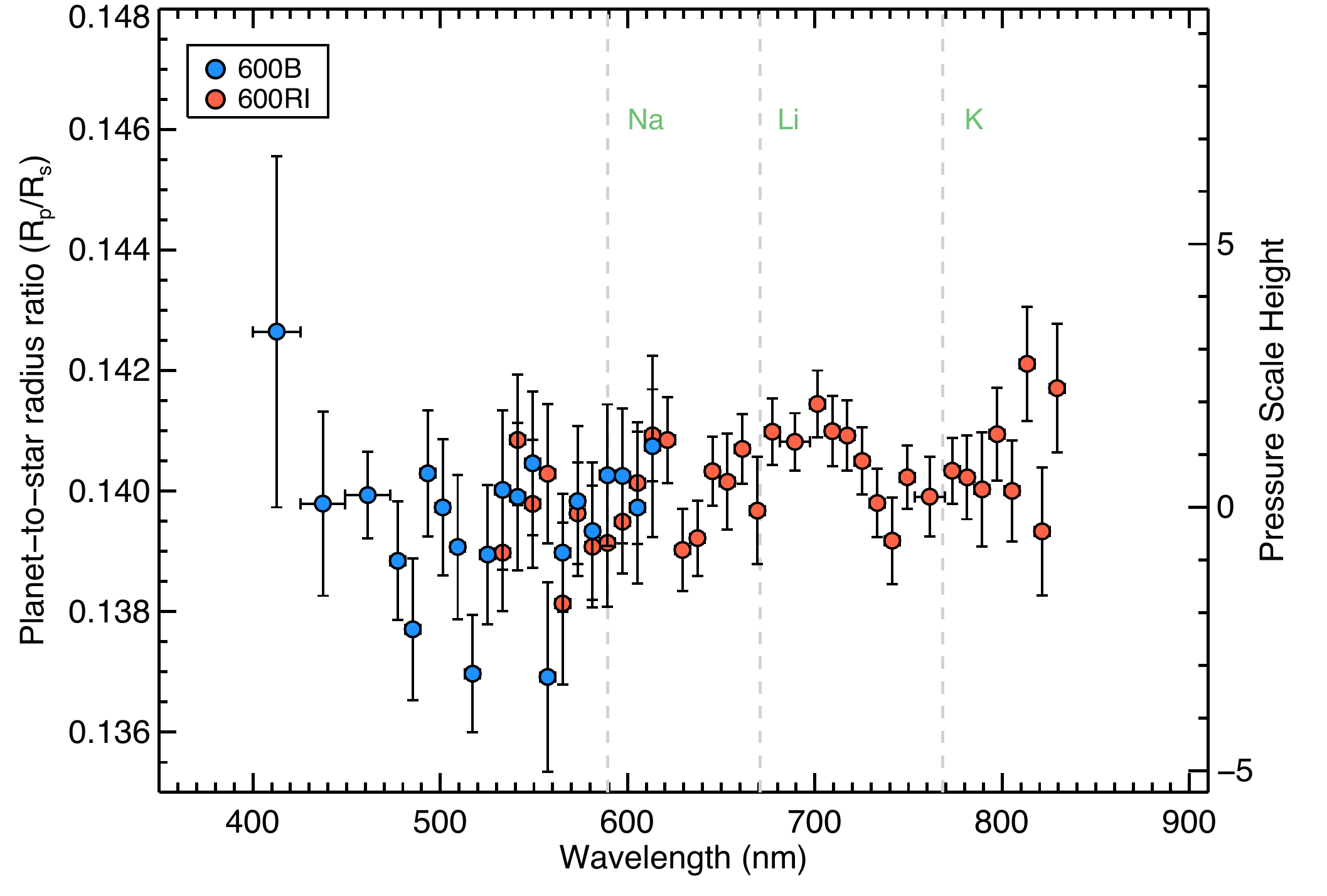}
\caption{{\rm{Indicated are the VLT FORS2 relative planet-to-star radius measurements from grism 600B (blue dots) and 600RI (red dots) along with the $1\sigma$ uncertainties.}} \label{fig:blud_red}}
\end{figure}

\begin{figure*}
\centering
\plotone{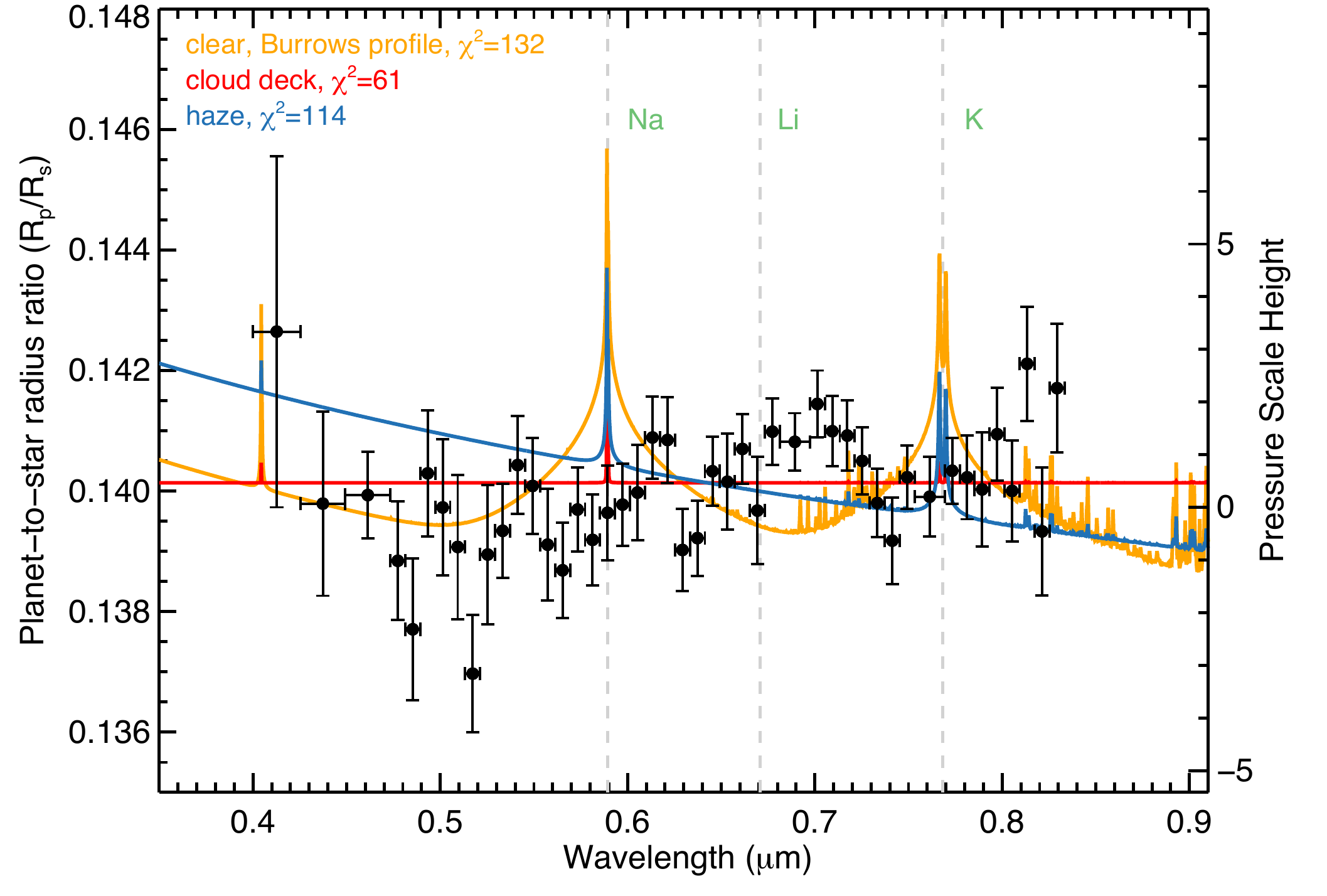}
\caption{Shown is the VLT transmission spectrum (black dots with $1\sigma$ vertical error bars; the horizontal bars indicate spectral bin widths) compared with a clear, cloudy and hazy one-dimensional forward atmospheric spectra at solar abundance (continuous lines). Line broadening shapes for Na and K have been calculated from the prescription detailed in \cite[][]{2000ApJ...531..438B}. Synthetic spectra with hazes or clouds (blue and red) predict much smaller and narrower absorption features.  \label{fig:fwd_mdls}}
\end{figure*}

The measured transmission spectrum of WASP-110b is presented in {\rm{Figure\,\ref{fig:blud_red} and \ref{fig:fwd_mdls}. To combine the transmission spectra from the blue and red grism observations, we computed the weighted mean of both datasets within the overlapping wavelength region. We find a marginally significant $\sim1.5\sigma$
radius difference from the two observations of $\Delta R_{p}/R_{\ast}=(33\pm22)\times10^{-4}$. This corresponds to a depth variation of $\sim11$\,ppm, which is within the $0.5\pm0.1\%$ stellar variation of the ASAS-SN photometry. The blue and red pairs of radius measurements show an excellent agreement within the overlapping region.}}

The spectrum is marginally flat without an increased absorption from sodium or potassium, which are the main absorbers expected in the optical transmission spectra of irradiated gas giants at $T_{{eq}}\sim1100$\,K similar to WASP-110b. An increased absorption is present in the wavelength range from $\sim0.63$ to $\sim0.75\mu$m, spanning $\sim3$ pressure scale heights. To interpret the spectrum we turn to model comparison and retrieval analysis.

\begin{deluxetable}{cccc}
\tablenum{4}
\tablecaption{Transmission spectrum of WASP-110b\label{tab:trspectab}}
\tablewidth{0pt}
\tablehead{
\colhead{Wavelength, \AA} & \colhead{$R_{p}/R_{\ast}$} & \colhead{$u_1$} & \colhead{$u_2$} 
}
\startdata
$4000-4253$ & $ 0.14265\pm 0.00291$ & $ 0.882\pm 0.030$ &  0.042\\
$4253-4493$ & $ 0.13979\pm 0.00153$ & $ 0.819\pm 0.018$ &  0.104\\
$4493-4733$ & $ 0.13993\pm 0.00072$ & $ 0.745\pm 0.011$ &  0.150\\
$4733-4813$ & $ 0.13884\pm 0.00099$ & $ 0.721\pm 0.018$ &  0.182\\
$4813-4893$ & $ 0.13771\pm 0.00118$ & $ 0.679\pm 0.022$ &  0.251\\
$4893-4973$ & $ 0.14029\pm 0.00104$ & $ 0.660\pm 0.021$ &  0.216\\
$4973-5053$ & $ 0.13973\pm 0.00113$ & $ 0.673\pm 0.019$ &  0.197\\
$5053-5133$ & $ 0.13907\pm 0.00120$ & $ 0.696\pm 0.021$ &  0.163\\
$5133-5213$ & $ 0.13697\pm 0.00097$ & $ 0.672\pm 0.021$ &  0.188\\
$5213-5293$ & $ 0.13895\pm 0.00115$ & $ 0.597\pm 0.022$ &  0.256\\
$5293-5373$ & $ 0.13934\pm 0.00078$ & $ 0.599\pm 0.023$ &  0.248\\
$5373-5453$ & $ 0.14043\pm 0.00081$ & $ 0.574\pm 0.021$ &  0.262\\
$5453-5533$ & $ 0.14009\pm 0.00079$ & $ 0.610\pm 0.021$ &  0.236\\
$5533-5613$ & $ 0.13911\pm 0.00093$ & $ 0.610\pm 0.022$ &  0.246\\
$5613-5693$ & $ 0.13869\pm 0.00079$ & $ 0.570\pm 0.021$ &  0.260\\
$5693-5773$ & $ 0.13969\pm 0.00070$ & $ 0.536\pm 0.026$ &  0.285\\
$5773-5853$ & $ 0.13919\pm 0.00076$ & $ 0.542\pm 0.022$ &  0.267\\
$5853-5933$ & $ 0.13964\pm 0.00078$ & $ 0.520\pm 0.026$ &  0.288\\
$5933-6013$ & $ 0.13978\pm 0.00068$ & $ 0.485\pm 0.023$ &  0.294\\
$6013-6093$ & $ 0.13998\pm 0.00079$ & $ 0.502\pm 0.027$ &  0.290\\
$6093-6173$ & $ 0.14089\pm 0.00068$ & $ 0.493\pm 0.027$ &  0.283\\
$6173-6253$ & $ 0.14085\pm 0.00071$ & $ 0.415\pm 0.020$ &  0.303\\
$6253-6333$ & $ 0.13902\pm 0.00068$ & $ 0.428\pm 0.019$ &  0.296\\
$6333-6413$ & $ 0.13922\pm 0.00062$ & $ 0.405\pm 0.019$ &  0.307\\
$6413-6493$ & $ 0.14033\pm 0.00057$ & $ 0.360\pm 0.019$ &  0.308\\
$6493-6573$ & $ 0.14015\pm 0.00080$ & $ 0.363\pm 0.019$ &  0.334\\
$6573-6653$ & $ 0.14070\pm 0.00057$ & $ 0.369\pm 0.018$ &  0.321\\
$6653-6733$ & $ 0.13968\pm 0.00089$ & $ 0.394\pm 0.023$ &  0.309\\
$6733-6813$ & $ 0.14099\pm 0.00055$ & $ 0.353\pm 0.020$ &  0.303\\
$6813-6973$ & $ 0.14082\pm 0.00048$ & $ 0.346\pm 0.016$ &  0.309\\
$6973-7053$ & $ 0.14145\pm 0.00055$ & $ 0.290\pm 0.020$ &  0.315\\
$7053-7133$ & $ 0.14099\pm 0.00058$ & $ 0.310\pm 0.024$ &  0.309\\
$7133-7213$ & $ 0.14092\pm 0.00058$ & $ 0.302\pm 0.024$ &  0.303\\
$7213-7293$ & $ 0.14050\pm 0.00056$ & $ 0.312\pm 0.021$ &  0.315\\
$7293-7373$ & $ 0.13980\pm 0.00056$ & $ 0.309\pm 0.024$ &  0.316\\
$7373-7453$ & $ 0.13918\pm 0.00072$ & $ 0.336\pm 0.026$ &  0.310\\
$7453-7533$ & $ 0.14023\pm 0.00052$ & $ 0.303\pm 0.021$ &  0.316\\
$7533-7693$ & $ 0.13991\pm 0.00066$ & $ 0.297\pm 0.020$ &  0.319\\
$7693-7773$ & $ 0.14034\pm 0.00055$ & $ 0.276\pm 0.024$ &  0.313\\
$7773-7853$ & $ 0.14023\pm 0.00069$ & $ 0.289\pm 0.025$ &  0.321\\
$7853-7933$ & $ 0.14003\pm 0.00095$ & $ 0.257\pm 0.027$ &  0.317\\
$7933-8013$ & $ 0.14094\pm 0.00078$ & $ 0.271\pm 0.026$ &  0.314\\
$8013-8093$ & $ 0.14000\pm 0.00084$ & $ 0.291\pm 0.022$ &  0.312\\
$8093-8173$ & $ 0.14211\pm 0.00095$ & $ 0.272\pm 0.026$ &  0.317\\
$8173-8253$ & $ 0.13933\pm 0.00106$ & $ 0.247\pm 0.025$ &  0.313\\
$8253-8333$ & $ 0.14171\pm 0.00107$ & $ 0.278\pm 0.025$ &  0.313\\
\enddata
\end{deluxetable}

\section{Discussion}\label{sec:discussion}
\subsection{Comparison to forward models}\label{sec:forward_models}
We compared the transmission spectrum with clear, cloudy and hazy synthetic spectra with solar abundances from the model presented in \cite{fortney08, 2010ApJ...709.1396F}. These spectra include a self-consistent treatment of radiative transfer and chemical equilibrium of neutral and ionic species. Chemical mixing ratios and opacities were computed assuming solar metallicity and local chemical equilibrium, accounting for condensation and thermal ionization but not photoionization \citep{lodders99, lodders02a, freedman08, 2014ApJS..214...25F}.

Similar to our previous studies, the cloudy and hazy models were computed using a simplified treatment of the scattering and absorption to simulate the effect of small particle haze aerosols and large particle cloud condensates at optical and near-infrared wavelengths. In the case of haze, Rayleigh scattering opacity $(\sigma = \sigma_0(\lambda/\lambda_0)^{-4})$ has been assumed with a cross-section which was $1,000\times$ the cross-section of molecular hydrogen gas $(\sigma_0 = 5.31 \times 10^{-27}$ cm$^2$ at $\lambda_0=3,500$\,\AA~\citep{lodders99}. To account for the effects of a flat cloud deck, we included a wavelength-independent cross-section, which was a factor of $100\times$ the cross-section of molecular hydrogen gas at $\lambda_0 = 3,500$\AA.

We obtained the average values of the models within the wavelength bins of the observed transmission spectrum and fitted these theoretical values to the data with a single parameter accounting for their vertical offset. The $\chi^2$ and Bayesian Information Criterion (BIC) statistic quantities were computed for each synthetic spectrum with the number of degrees of freedom for each model determined by $\nu = N - m$, where $N$ is the number of measurements and $m$ is the number of free parameters in the fit. We find the cloudy spectra to best describe our observations ($\chi^2=61$) followed by the haze and cloud-free simulated spectra (Figure\,\ref{fig:fwd_mdls})

\subsection{Retrieval analysis}\label{sec:retrieval}

\begin{figure}[ht!]
    \centering
    \includegraphics[width=\columnwidth]{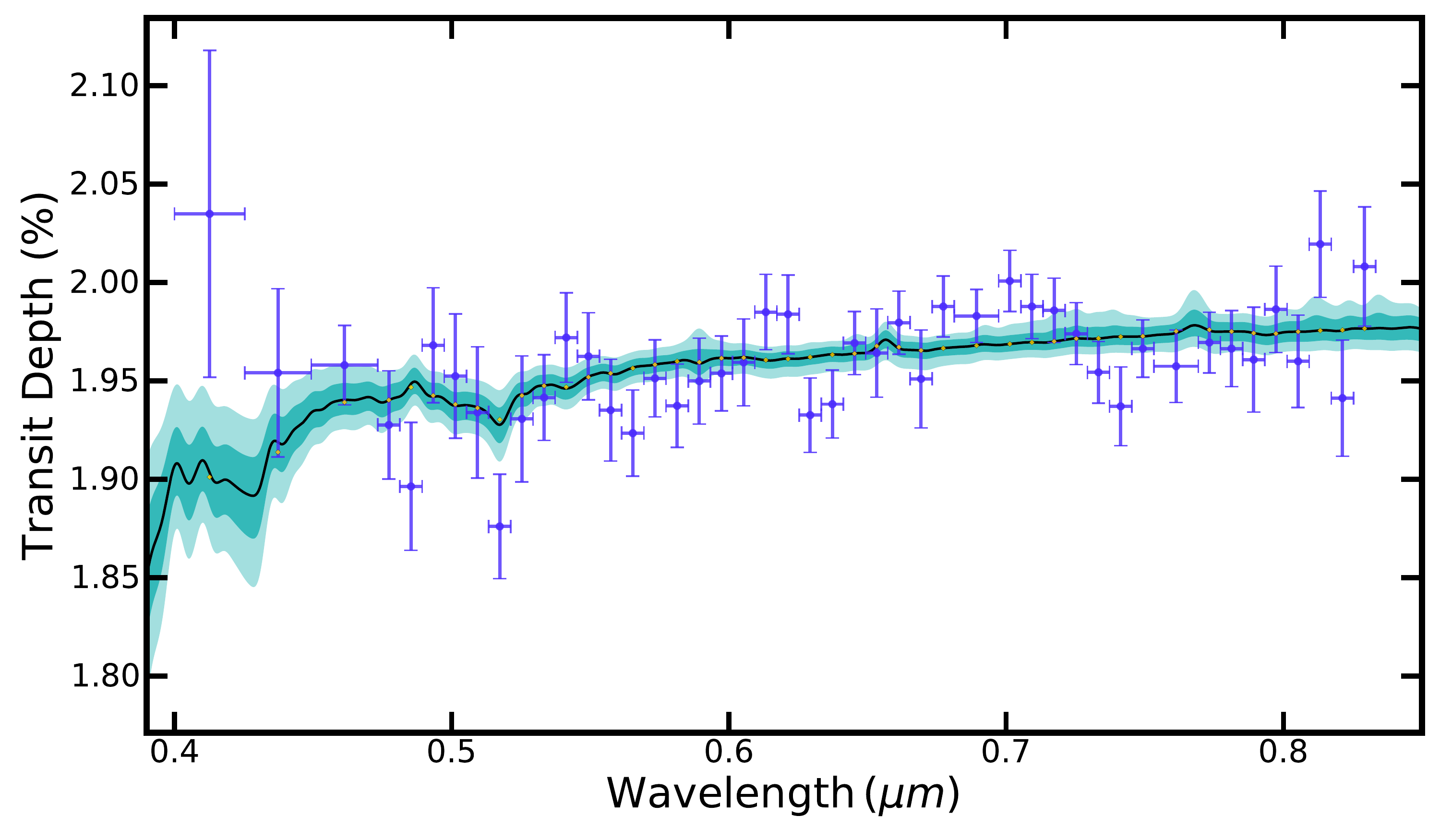}
    \caption{Our VLT FORS2 data (blue errorbars) and the retrieved median spectrum (black) and corresponding 1- and 2-$\sigma$ contours (dark and light turquoise) from our AURA retrieval. The median spectral fit has an increasing transit depth with wavelength, due to faculae on the unobstructed stellar surface. It does not have any significant absorption features from chemical species.}
    \label{fig:AURA_spectrum}
\end{figure}

\begin{figure}[ht!]
    \centering
    \includegraphics[width=\columnwidth]{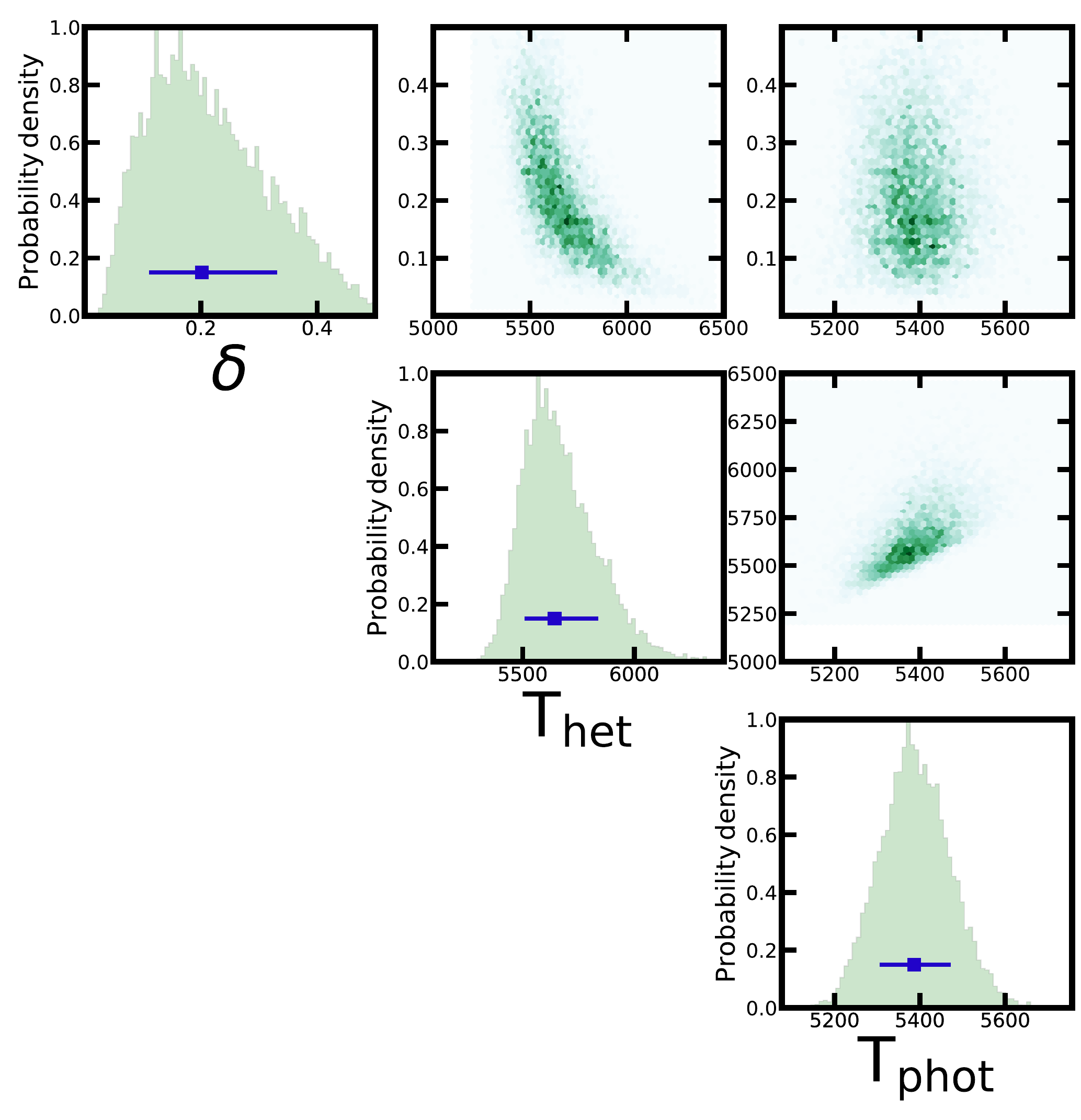}
    \caption{The posterior probability distributions for the three stellar heterogeneity parameters used in our AURA retrieval. Our retrieval successfully constrains all three. We retrieved a T$_{\mathrm{het}}$ estimate that is higher than both the literature T$_{\mathrm{eff}}$ and retrieved T$_{\mathrm{phot}}$ values, indicating that hotter faculae are the dominant form of stellar heterogeneity.}
    \label{fig:AURA_heterogeneity_posterior}
\end{figure}

We additionally considered inverse methods to constrain the atmospheric properties of WASP-110b via atmospheric retrievals using the AURA code \citep{Pinhas2018}. AURA computes model transmission spectra using line-by-line radiative transfer in a plane parallel atmosphere in hydrostatic equilibrium. The forward model-generating component is interfaced with the PyMultiNest Nested Sampling package \citep{Buchner2014}, which allows for Bayesian parameter estimation and model comparison. We describe the terminator pressure-temperature profile using the 6-parameter prescription of \citet{Madhusudhan2009}. Our model atmosphere includes opacities arising from H$_2$-H$_2$ and H$_2$-He collision-induced absorption, as well as absorption from H$_2$O \citep{Rothman2010}, Na and K \citep{Welbanks2019b}, the latter two accounting for the effects of pressure broadening. We avoid including high temperature species in our atmospheric model (e.g. TiO, VO and other metal hydrides/oxides, ions, etc), given the relatively low equilibrium temperature of WASP-110~b. Inhomogeneous coverage of clouds and hazes is incorporated in our model using the prescription of \citet{Pinhas2018}, treating clouds as grey opacity below the cloud deck and hazes as a modification to Rayleigh scattering above the cloud deck.

AURA also incorporates the effects of stellar heterogeneity, being the first retrieval code demonstrated to successfully include such effects \citep{Pinhas2018}. The stellar heterogeneities are described using 3 parameters: $\delta$, the fraction of the projected stellar disc covered by cooler starspots/hotter faculae, T$_{\mathrm{het}}$ the temperature of the starspots/faculae and T$_{\mathrm{phot}}$ the temperature of the immaculate photosphere. We construct the spectral contributions of the two stellar surface components by interpolating through a grid of PHOENIX spectra \citep{Husser2013}, fixing the stellar metallicity and gravity to literature values. Hotter faculae/cooler starspots result in a larger/smaller apparent stellar radius towards the blue end of the spectrum, which in turn results in lower/higher observed transit depths at shorter wavelengths.

We find that our retrieval produces a good fit to the data, as show in Figure \ref{fig:AURA_spectrum}. We successfully obtain constraints for the three stellar heterogeneity parameters, finding $\delta = 0.20^{+0.13}_{-0.09}$, T$_{\mathrm{het}} = 5643^{+195}_{-134}$\,K and T$_{\mathrm{phot}} = 5387^{+85}_{-81}$\,K. We show the retrieved posterior distribution for these three parameters in Figure \ref{fig:AURA_heterogeneity_posterior}. Our retrieved T$_{\mathrm{het}}$ value is higher than both the retrieved T$_{\mathrm{phot}}$ and literature T$_{\mathrm{eff}}$ values, indicating the presence of hotter faculae. We determined the detection significance (D.S.) for stellar heterogeneity by carrying out a Bayesian model comparison with the findings of a second retrieval that excludes the effects of stellar heterogeneity. We find a D.S. of 3.1~$\sigma$, indicating a strong detection. {\rm This stems from the overall trend in the data of increasing transit depths at longer wavelengths, which are attributed to unocculted stellar faculae rather than opacity sources in the planetary atmosphere.}

{\rm We additionally consider separate retrievals for the grism 600B and 600RI datasets, without applying any offsets to either. We find that the constraints on the cloud and heterogeneity parameters are governed primarily by the 600RI dataset, with the 600B dataset providing only weak constraints. The constraints from both datasets are consistent with each other and with the joint dataset to within the 1-$\sigma$ uncertainties. We, therefore, find that the two datasets can be combined for a joint retrieval with a single set of heterogeneity parameters without significant loss of information.}

Our retrieval is unable to detect any of the chemical species considered, including H$_2$O. We additionally obtain weak constraints for a high-altitude cloud deck pressure, $\log ( P_{\mathrm{cloud}}) = -3.4^{+1.8}_{-1.3}$ (in bar), while also finding that our retrieval shows a preference for near-full cloud coverage fractions. As before, we conducted a third retrieval, this time excluding clouds and hazes to determine their D.S.. We find that our retrievals detect clouds and hazes with a 2.1~$\sigma$ significance, suggesting only a tentative detection. {\rm Clouds are only tentatively detected because a featureless spectrum can be caused by either clouds or low abundances of alkali metals.} The overall posterior distribution is shown in Figure \ref{fig:AURA_full_posterior}.

{\rm We also conduct further retrievals to explore how more elaborate model considerations affect our findings. In one case, we additionally consider several chemical species with significant optical opacities, such as TiO, VO and AlO \citep{fortney08,gandhi2019}, even though they are not expected in thermochemical equilibrium at the low atmospheric temperatures of WASP-110~b. While we find nominal indications of these species, none has a D.S. greater than 2 $\sigma$ to constituting a strong detection. We also run retrievals with an offset applied to either the 600RI or 600B datasets as an additional retrieved parameter. In this case, the D.S. of the heterogeneity is reduced to $\sim$ 1-2$\sigma$, depending on which of the two datasets is being offset, while clouds and hazes are detected at a D.S. of $\sim$2 $\sigma$.}

\section{A comparison to a similar irradiated exoplanet}\label{sec:clear_cloudy_hazy}
Atmospheric temperature and gravity are two fundamental properties considered to play a definitive role in the formation of clouds and hazes. With a growing sample of irradiated gas giants characterized in transmission, it becomes imperative to identify similar exoplanets and compare their spectra and physical characteristics. Similar properties have the potential to reveal transitions of atmospheric types and ultimately identify the factors determining why some planets appear cloud-free and others cloudy or hazy. The physical properties of WASP-110b and WASP-6b are nearly identical, which also holds for their host stars (Table\,\ref{tab:clear_cloudy_hazy_tab}). With a surface gravity of $\sim8\,$\,ms$^{-2}$ and an atmospheric temperature estimated to $\lesssim1200$\,K, the transmission spectra for both atmospheres are predicted to be dominated by the pressure-broadened absorption features of sodium and potassium. This partially holds for WASP-6b, where observations of {\it{HST}} and VLT reveal absorption from sodium, potassium and water and an optical to infrared slope spanning $\sim8$ scale heights {\rm{\citep{carter20}}}. {\rm{However, with a lack of the theoretically predicted pressure broadened sodium and potassium absorption features, the spectrum of WASP-110b drastically differs from the spectrum of WASP-6b and this difference may likely be attributed to the impact of clouds.}} 

Despite the remarkable similarities in their physical properties, the significant difference in the optical spectra for both planets indicate that additional factors are responsible for determining the presence of clouds and hazes. One such factor is the temperature-pressure profile at the day-night terminator, which can be specific for the west and east limbs. A step toward capturing these effects consistently would involve three-dimensional (3D) simulations that include clouds  \citep[e.g.][]{2018A&A...615A..97L}, chemical kinetics \citep[e.g.][]{2020A&A...636A..68D}, etc., and should start with virtually the same inputs. Given that planetary atmospheres are complex and nonlinear systems, subtle variations from planet to planet, including measured system properties (mass, radius, metallicity), higher-order properties (gravitational and magnetic fields) and at levels below the precisions achieved by transmission spectroscopy observations reported in Table\,\ref{tab:trspectab} can result in significantly different atmosphere signals. Furthermore, planetary atmospheres could be time variable, as seen in general circulation models (GCMs) and our own solar system, e.g. oscillations in Earth's history into Snowball states, global storms on Mars triggered by subtle effects etc. Future observations, including with {\it{JWST}} instruments (e.g. MIRI) are expected constrain the planet temperature and shed more light on the composition of the clouds and hazes of exoplanet atmospheres \citep{2015A&A...573A.122W}. 


\begin{deluxetable}{lcc}[t]
\tablenum{5}
\tablecaption{Physical properties\label{tab:clear_cloudy_hazy_tab}}
\tablewidth{0pt}
\tablehead{
\colhead{Property}          & WASP-110b$^{\rm{A}}$ & WASP-6b$^{\rm{B}}$
}
\startdata
Atm. type                   & $cloudy$                & $hazy$             \\
Constituent                 & -                       & Na, K, H$_2$O      \\
H, km                       & 536                     & 533                \\
$T_{\rm{eq}}$, K            & $1134\pm33$             & $ 1184\pm16$       \\
g, m s$^{-2}$               & $7.60\pm1$              & $ 7.96\pm0.30$      \\
M$_p$, M$_{J}$              & $0.510\pm0.064$         & $ 0.485\pm0.027$    \\
R$_p$, M$_{J}$              & $ 1.238\pm0.056$        & $ 1.230\pm0.035$   \\
P, day                      & 3.78                    & 3.36                \\
a, au                       & 0.05                    & 0.04                      \\
e                           & $0^{+0.61}_{-0}$        & $0.054^{+0.018}_{-0.014}$   \\
$T_{\rm{eff}}$, K           & $5360\pm130$            & $5375\pm65$          \\
$\log{g}$, cgs              & $ 4.498\pm0.022$        & $4.487\pm0.017$      \\
$[$Fe/H$]$.                 & $-0.06\pm0.10$          & $-0.15\pm0.09$     \\
M$_\ast$, M$_\odot$         & $0.892\pm0.072$         & $0.836\pm0.063$      \\
R$_\ast$, M$_\odot$         & $0.881\pm0.035$         & $0.864\pm0.024$    \\
$\log{R'_{\rm HK}}$         & $-4.9^{+0.09}_{-0.24}$  & $-4.511\pm0.037$       \\
\enddata
\tablecomments{\,A: \cite{Anderson2014}; B: \cite{Gillon09}.}
\end{deluxetable}

While WASP-110 and WASP-6 are two alike stars with respect to their fundamental physical properties, {\mbox{WASP-6}} exhibits a higher level of chromospheric activity, as revealed by the higher $\log{R'_{\rm HK}}$ index. Active stars undergoing magnetic cycles modulate cosmic ray fluxes, which can exceed the rate of galactic cosmic rays in the vicinity of the star. The formation of clouds enhances in the presence of a nucleation particle, including liquid or solid particles, such as droplets or grains as they act as nuclei for further condensation \citep[][]{2015plsc.book.....D}. Galactic and solar cosmic rays, mainly protons, reaching Earth's atmosphere carry enough energy to ionize volatile compounds causing them to condense into droplets, or aerosols and enhance cloud formation \citep[][]{1998PhRvL..81.5027S, 2007SGeo...28..333K, 2009GeoRL..3615101S, 2011Natur.476..429K}. Due to the higher activity level of the host star, it would be natural to assume that WASP-6b experiences a higher flux of cosmic rays and more nucleation events that promote cloud formation in its atmosphere. Planetary magnetic fields are considered to play a shielding role from high energy particles and reduce the formation of nucleation particles and clouds. Could WASP-6b possess a magnetic field that shields its atmosphere and reduces the flux of cosmic rays while WASP-110b's magnetic field is much weaker? Understanding the global picture of exoplanet atmospheres requires the full asset of planetary physical properties including the planetary magnetic field, which are expected to be probed with the next generation of instruments and would help elucidate the picture toward these two exoplanet atmospheres.

\section{Summary}\label{sec:Summary}
We have conducted a comparative forward and retrieval analysis of the 0.4 to 0.833 $\mu$m transmission spectrum of WASP-110b, obtained using VLT FORS2 transit spectroscopy (Section\,\ref{sec:vlt_obs}). We have improved the planet orbital ephemerides and found no departures from the predicted transit times from analysis of six transits observed by the TESS space-based photometer (Section\,\ref{sec:tess_obs}). We determined the host star exhibits low to moderate activity ($\log{R'_{\rm HK}}=-4.9^{+0.09}_{-0.24}$) using optical high-resolution MPG/ESO2.2 FEROS spectroscopy (Section\,\ref{sec:activity}). 

We have compared the VLT transmission spectrum to a set of clear, cloudy and hazy forward hot Jupiter atmosphere spectra (Section\,\ref{sec:forward_models}). This analysis indicates a cloudy atmosphere. We have performed retrieval analysis with the AURA {\rm retrieval framework}, which includes contamination from stellar heterogeneity (Section\,\ref{sec:retrieval}). The fiducial model from the retrieval analyses is an excellent fit to the transmission spectrum. The retrieval results point toward a scenario of unocculted faculae with temperature $\sim200$K hotter than the stellar photosphere and fractional coverage of $\sim20\%$ in combination with a cloud deck as the explanation {\rm for} the observed VLT spectrum.

\acknowledgments{Based on observations made with ESO Telescopes at the La Silla Paranal Observatory under programme ID 199.C-0467 and 099.A-9010. The authors are grateful to Dr.~David Anderson and Dr.~David Brown for sharing the follow-up light curve with us. GM acknowledges the financial support from the National Science Centre, Poland through grant no. 2016/23/B/ST9/00579. The authors are grateful to N\'{e}stor Espinoza, Jayesh Goyal, Ryan MacDonald and Joanna Barstow for the valuable discussions, which helped to improve the manuscript. {\rm{The authors are grateful to the anonymous reviewer for their constructive comments on the manuscript.}} This work has made use of data from the European Space Agency (ESA) mission {\it Gaia} (\url{https://www.cosmos.esa.int/gaia}), processed by the {\it Gaia} Data Processing and Analysis Consortium (DPAC, \url{https://www.cosmos.esa.int/web/gaia/dpac/consortium}). Funding for the DPAC has been provided by national institutions, in particular the institutions participating in the {\it Gaia} Multilateral Agreement. This work made use of the Python Gaussian process library George. NM acknowledges funding from the UKRI Future Leaders Scheme (MR/T040866/1), Science and Technology Facilities Council Consolidated Grant (ST/R000395/1) and Leverhulme Trust research project grant (RPG-2020-82).}

\software{george  \citep{Foreman-Mackey15}, TESSCut tool \citep{2019ascl.soft05007B},  Lightkurve v1.9 \citep{2018ascl.soft12013L}, TAP, \citet{2012AdAst2012E..30G}, IRAF \citep{1986SPIE..627..733T, 1993ASPC...52..173T}, hjd2bjd.html \citep{2010PASP..122..935E}, MPFIT  \citep[][]{markwardt09}, Astropy \citep{astropy:2013, astropy:2018}, AURA \citep{Pinhas2018}, PyMultiNest \citep{Buchner2014}}


%

\facilities{VLT(FORS2), TESS, MPG/ESO2.2(FEROS)}

\bibliography{sample63}{}

\begin{thebibliography}{}
\expandafter\ifx\csname natexlab\endcsname\relax\def\natexlab#1{#1}\fi
\providecommand{\url}[1]{\href{#1}{#1}}
\providecommand{\dodoi}[1]{doi:~\href{http://doi.org/#1}{\nolinkurl{#1}}}
\providecommand{\doeprint}[1]{\href{http://ascl.net/#1}{\nolinkurl{http://ascl.net/#1}}}
\providecommand{\doarXiv}[1]{\href{https://arxiv.org/abs/#1}{\nolinkurl{https://arxiv.org/abs/#1}}}

\bibitem[{{Akaike}(1974)}]{akaike74}
{Akaike}, H. 1974, IEEE Transactions on Automatic Control, 19, 716

\bibitem[{{Anderson} {et~al.}(2014{\natexlab{a}}){Anderson}, {Brown}, {Collier
  Cameron}, {Delrez}, {Fumel}, {Gillon}, {Hellier}, {Jehin}, {Lendl}, {Maxted},
  {Neveu-VanMalle}, {Pepe}, {Pollacco}, {Queloz}, {Rojo}, {Segransan},
  {Serenelli}, {Smalley}, {Smith}, {Southworth}, {Triaud}, {Turner}, {Udry}, \&
  {West}}]{Anderson2014}
{Anderson}, D.~R., {Brown}, D.~J.~A., {Collier Cameron}, A., {et~al.}
  2014{\natexlab{a}}, arXiv e-prints, arXiv:1410.3449.
\newblock \doarXiv{1410.3449}

\bibitem[{{Anderson} {et~al.}(2014{\natexlab{b}}){Anderson}, {Brown}, {Collier
  Cameron}, {Delrez}, {Fumel}, {Gillon}, {Hellier}, {Jehin}, {Lendl}, {Maxted},
  {Neveu-VanMalle}, {Pepe}, {Pollacco}, {Queloz}, {Rojo}, {Segransan},
  {Serenelli}, {Smalley}, {Smith}, {Southworth}, {Triaud}, {Turner}, {Udry}, \&
  {West}}]{2014arXiv1410.3449A}
---. 2014{\natexlab{b}}, arXiv e-prints, arXiv:1410.3449.
\newblock \doarXiv{1410.3449}

\bibitem[{{Appenzeller} {et~al.}(1998){Appenzeller}, {Fricke}, {F{\"u}rtig},
  {G{\"a}ssler}, {H{\"a}fner}, {Harke}, {Hess}, {Hummel}, {J{\"u}rgens},
  {Kudritzki}, {Mantel}, {Meisl}, {Muschielok}, {Nicklas}, {Rupprecht},
  {Seifert}, {Stahl}, {Szeifert}, \& {Tarantik}}]{Appenzeller1998}
{Appenzeller}, I., {Fricke}, K., {F{\"u}rtig}, W., {et~al.} 1998, The
  Messenger, 94, 1

\bibitem[{{Astropy Collaboration} {et~al.}(2013){Astropy Collaboration},
  {Robitaille}, {Tollerud}, {Greenfield}, {Droettboom}, {Bray}, {Aldcroft},
  {Davis}, {Ginsburg}, {Price-Whelan}, {Kerzendorf}, {Conley}, {Crighton},
  {Barbary}, {Muna}, {Ferguson}, {Grollier}, {Parikh}, {Nair}, {Unther},
  {Deil}, {Woillez}, {Conseil}, {Kramer}, {Turner}, {Singer}, {Fox}, {Weaver},
  {Zabalza}, {Edwards}, {Azalee Bostroem}, {Burke}, {Casey}, {Crawford},
  {Dencheva}, {Ely}, {Jenness}, {Labrie}, {Lim}, {Pierfederici}, {Pontzen},
  {Ptak}, {Refsdal}, {Servillat}, \& {Streicher}}]{astropy:2013}
{Astropy Collaboration}, {Robitaille}, T.~P., {Tollerud}, E.~J., {et~al.} 2013,
  \aap, 558, A33, \dodoi{10.1051/0004-6361/201322068}

\bibitem[{{Astropy Collaboration} {et~al.}(2018){Astropy Collaboration},
  {Price-Whelan}, {Sip{\H{o}}cz}, {G{\"u}nther}, {Lim}, {Crawford}, {Conseil},
  {Shupe}, {Craig}, {Dencheva}, {Ginsburg}, {Vand erPlas}, {Bradley},
  {P{\'e}rez-Su{\'a}rez}, {de Val-Borro}, {Aldcroft}, {Cruz}, {Robitaille},
  {Tollerud}, {Ardelean}, {Babej}, {Bach}, {Bachetti}, {Bakanov}, {Bamford},
  {Barentsen}, {Barmby}, {Baumbach}, {Berry}, {Biscani}, {Boquien}, {Bostroem},
  {Bouma}, {Brammer}, {Bray}, {Breytenbach}, {Buddelmeijer}, {Burke},
  {Calderone}, {Cano Rodr{\'\i}guez}, {Cara}, {Cardoso}, {Cheedella}, {Copin},
  {Corrales}, {Crichton}, {D'Avella}, {Deil}, {Depagne}, {Dietrich}, {Donath},
  {Droettboom}, {Earl}, {Erben}, {Fabbro}, {Ferreira}, {Finethy}, {Fox},
  {Garrison}, {Gibbons}, {Goldstein}, {Gommers}, {Greco}, {Greenfield},
  {Groener}, {Grollier}, {Hagen}, {Hirst}, {Homeier}, {Horton}, {Hosseinzadeh},
  {Hu}, {Hunkeler}, {Ivezi{\'c}}, {Jain}, {Jenness}, {Kanarek}, {Kendrew},
  {Kern}, {Kerzendorf}, {Khvalko}, {King}, {Kirkby}, {Kulkarni}, {Kumar},
  {Lee}, {Lenz}, {Littlefair}, {Ma}, {Macleod}, {Mastropietro}, {McCully},
  {Montagnac}, {Morris}, {Mueller}, {Mumford}, {Muna}, {Murphy}, {Nelson},
  {Nguyen}, {Ninan}, {N{\"o}the}, {Ogaz}, {Oh}, {Parejko}, {Parley}, {Pascual},
  {Patil}, {Patil}, {Plunkett}, {Prochaska}, {Rastogi}, {Reddy Janga},
  {Sabater}, {Sakurikar}, {Seifert}, {Sherbert}, {Sherwood-Taylor}, {Shih},
  {Sick}, {Silbiger}, {Singanamalla}, {Singer}, {Sladen}, {Sooley},
  {Sornarajah}, {Streicher}, {Teuben}, {Thomas}, {Tremblay}, {Turner},
  {Terr{\'o}n}, {van Kerkwijk}, {de la Vega}, {Watkins}, {Weaver}, {Whitmore},
  {Woillez}, {Zabalza}, \& {Astropy Contributors}}]{astropy:2018}
{Astropy Collaboration}, {Price-Whelan}, A.~M., {Sip{\H{o}}cz}, B.~M., {et~al.}
  2018, \aj, 156, 123, \dodoi{10.3847/1538-3881/aabc4f}

\bibitem[{{Brasseur} {et~al.}(2019){Brasseur}, {Phillip}, {Fleming},
  {Mullally}, \& {White}}]{2019ascl.soft05007B}
{Brasseur}, C.~E., {Phillip}, C., {Fleming}, S.~W., {Mullally}, S.~E., \&
  {White}, R.~L. 2019, {Astrocut: Tools for creating cutouts of TESS images}.
\newblock \doeprint{1905.007}

\bibitem[{{Buchner} {et~al.}(2014){Buchner}, {Georgakakis}, {Nandra}, {Hsu},
  {Rangel}, {Brightman}, {Merloni}, {Salvato}, {Donley}, \&
  {Kocevski}}]{Buchner2014}
{Buchner}, J., {Georgakakis}, A., {Nandra}, K., {et~al.} 2014, \aap, 564, A125,
  \dodoi{10.1051/0004-6361/201322971}

\bibitem[{{Burrows} {et~al.}(2000){Burrows}, {Marley}, \&
  {Sharp}}]{2000ApJ...531..438B}
{Burrows}, A., {Marley}, M.~S., \& {Sharp}, C.~M. 2000, \apj, 531, 438,
  \dodoi{10.1086/308462}

\bibitem[{{Carter} {et~al.}(2020){Carter}, {Nikolov}, {Sing}, {Alam}, {Goyal},
  {Mikal-Evans}, {Wakeford}, {Henry}, {Morrell}, {L{\'o}pez-Morales},
  {Smalley}, {Lavvas}, {Barstow}, {Garc{\'\i}a Mu{\~n}oz}, {Gibson}, \&
  {Wilson}}]{carter20}
{Carter}, A.~L., {Nikolov}, N., {Sing}, D.~K., {et~al.} 2020, \mnras, 494,
  5449, \dodoi{10.1093/mnras/staa1078}

\bibitem[{{Claret} \& {Bloemen}(2011)}]{2011A&A...529A..75C}
{Claret}, A., \& {Bloemen}, S. 2011, \aap, 529, A75,
  \dodoi{10.1051/0004-6361/201116451}

\bibitem[{{de Pater} \& {Lissauer}(2015)}]{2015plsc.book.....D}
{de Pater}, I., \& {Lissauer}, J.~J. 2015, {Planetary Sciences}

\bibitem[{{Drummond} {et~al.}(2020){Drummond}, {H{\'e}brard}, {Mayne}, {Venot},
  {Ridgway}, {Changeat}, {Tsai}, {Manners}, {Tremblin}, {Abraham}, {Sing}, \&
  {Kohary}}]{2020A&A...636A..68D}
{Drummond}, B., {H{\'e}brard}, E., {Mayne}, N.~J., {et~al.} 2020, \aap, 636,
  A68, \dodoi{10.1051/0004-6361/201937153}

\bibitem[{{Eastman} {et~al.}(2010){Eastman}, {Siverd}, \&
  {Gaudi}}]{2010PASP..122..935E}
{Eastman}, J., {Siverd}, R., \& {Gaudi}, B.~S. 2010, \pasp, 122, 935,
  \dodoi{10.1086/655938}

\bibitem[{{Espinoza} \& {Jord{\'a}n}(2016)}]{espinoza2016}
{Espinoza}, N., \& {Jord{\'a}n}, A. 2016, \mnras, 457, 3573,
  \dodoi{10.1093/mnras/stw224}

\bibitem[{{Foreman-Mackey}(2015)}]{Foreman-Mackey15}
{Foreman-Mackey}, D. 2015, {George: Gaussian Process regression}, Astrophysics
  Source Code Library.
\newblock \doeprint{1511.015}

\bibitem[{{Foreman-Mackey} {et~al.}(2013){Foreman-Mackey}, {Hogg}, {Lang}, \&
  {Goodman}}]{Foreman13}
{Foreman-Mackey}, D., {Hogg}, D.~W., {Lang}, D., \& {Goodman}, J. 2013, \pasp,
  125, 306, \dodoi{10.1086/670067}

\bibitem[{{Fortney} {et~al.}(2008){Fortney}, {Lodders}, {Marley}, \&
  {Freedman}}]{fortney08}
{Fortney}, J.~J., {Lodders}, K., {Marley}, M.~S., \& {Freedman}, R.~S. 2008,
  \apj, 678, 1419, \dodoi{10.1086/528370}

\bibitem[{{Fortney} {et~al.}(2010){Fortney}, {Shabram}, {Showman}, {Lian},
  {Freedman}, {Marley}, \& {Lewis}}]{2010ApJ...709.1396F}
{Fortney}, J.~J., {Shabram}, M., {Showman}, A.~P., {et~al.} 2010, \apj, 709,
  1396, \dodoi{10.1088/0004-637X/709/2/1396}

\bibitem[{{Freedman} {et~al.}(2014){Freedman}, {Lustig-Yaeger}, {Fortney},
  {Lupu}, {Marley}, \& {Lodders}}]{2014ApJS..214...25F}
{Freedman}, R.~S., {Lustig-Yaeger}, J., {Fortney}, J.~J., {et~al.} 2014, \apjs,
  214, 25, \dodoi{10.1088/0067-0049/214/2/25}

\bibitem[{{Freedman} {et~al.}(2008){Freedman}, {Marley}, \&
  {Lodders}}]{freedman08}
{Freedman}, R.~S., {Marley}, M.~S., \& {Lodders}, K. 2008, \apjs, 174, 504,
  \dodoi{10.1086/521793}

\bibitem[{{Fulton} {et~al.}(2011){Fulton}, {Shporer}, {Winn}, {Holman},
  {P{\'a}l}, \& {Gazak}}]{2011AJ....142...84F}
{Fulton}, B.~J., {Shporer}, A., {Winn}, J.~N., {et~al.} 2011, \aj, 142, 84,
  \dodoi{10.1088/0004-6256/142/3/84}

\bibitem[{{Gaia Collaboration} {et~al.}(2018){Gaia Collaboration}, {Brown},
  {Vallenari}, {Prusti}, {de Bruijne}, {Babusiaux}, {Bailer-Jones}, {Biermann},
  {Evans}, {Eyer}, {Jansen}, {Jordi}, {Klioner}, {Lammers}, {Lindegren},
  {Luri}, {Mignard}, {Panem}, {Pourbaix}, {Randich}, {Sartoretti}, {Siddiqui},
  {Soubiran}, {van Leeuwen}, {Walton}, {Arenou}, {Bastian}, {Cropper},
  {Drimmel}, {Katz}, {Lattanzi}, {Bakker}, {Cacciari}, {Casta{\~n}eda},
  {Chaoul}, {Cheek}, {De Angeli}, {Fabricius}, {Guerra}, {Holl}, {Masana},
  {Messineo}, {Mowlavi}, {Nienartowicz}, {Panuzzo}, {Portell}, {Riello},
  {Seabroke}, {Tanga}, {Th{\'e}venin}, {Gracia-Abril}, {Comoretto},
  {Garcia-Reinaldos}, {Teyssier}, {Altmann}, {Andrae}, {Audard},
  {Bellas-Velidis}, {Benson}, {Berthier}, {Blomme}, {Burgess}, {Busso},
  {Carry}, {Cellino}, {Clementini}, {Clotet}, {Creevey}, {Davidson}, {De
  Ridder}, {Delchambre}, {Dell'Oro}, {Ducourant},
  {Fern{\'a}ndez-Hern{\'a}ndez}, {Fouesneau}, {Fr{\'e}mat}, {Galluccio},
  {Garc{\'\i}a-Torres}, {Gonz{\'a}lez-N{\'u}{\~n}ez}, {Gonz{\'a}lez-Vidal},
  {Gosset}, {Guy}, {Halbwachs}, {Hambly}, {Harrison}, {Hern{\'a}ndez},
  {Hestroffer}, {Hodgkin}, {Hutton}, {Jasniewicz}, {Jean-Antoine-Piccolo},
  {Jordan}, {Korn}, {Krone-Martins}, {Lanzafame}, {Lebzelter}, {L{\"o}ffler},
  {Manteiga}, {Marrese}, {Mart{\'\i}n-Fleitas}, {Moitinho}, {Mora}, {Muinonen},
  {Osinde}, {Pancino}, {Pauwels}, {Petit}, {Recio-Blanco}, {Richards},
  {Rimoldini}, {Robin}, {Sarro}, {Siopis}, {Smith}, {Sozzetti}, {S{\"u}veges},
  {Torra}, {van Reeven}, {Abbas}, {Abreu Aramburu}, {Accart}, {Aerts},
  {Altavilla}, {{\'A}lvarez}, {Alvarez}, {Alves}, {Anderson}, {Andrei},
  {Anglada Varela}, {Antiche}, {Antoja}, {Arcay}, {Astraatmadja}, {Bach},
  {Baker}, {Balaguer-N{\'u}{\~n}ez}, {Balm}, {Barache}, {Barata}, {Barbato},
  {Barblan}, {Barklem}, {Barrado}, {Barros}, {Barstow}, {Bartholom{\'e}
  Mu{\~n}oz}, {Bassilana}, {Becciani}, {Bellazzini}, {Berihuete}, {Bertone},
  {Bianchi}, {Bienaym{\'e}}, {Blanco-Cuaresma}, {Boch}, {Boeche}, {Bombrun},
  {Borrachero}, {Bossini}, {Bouquillon}, {Bourda}, {Bragaglia}, {Bramante},
  {Breddels}, {Bressan}, {Brouillet}, {Br{\"u}semeister}, {Brugaletta},
  {Bucciarelli}, {Burlacu}, {Busonero}, {Butkevich}, {Buzzi}, {Caffau},
  {Cancelliere}, {Cannizzaro}, {Cantat-Gaudin}, {Carballo}, {Carlucci},
  {Carrasco}, {Casamiquela}, {Castellani}, {Castro-Ginard}, {Charlot},
  {Chemin}, {Chiavassa}, {Cocozza}, {Costigan}, {Cowell}, {Crifo}, {Crosta},
  {Crowley}, {Cuypers}, {Dafonte}, {Damerdji}, {Dapergolas}, {David}, {David},
  {de Laverny}, {De Luise}, {De March}, {de Martino}, {de Souza}, {de Torres},
  {Debosscher}, {del Pozo}, {Delbo}, {Delgado}, {Delgado}, {Di Matteo},
  {Diakite}, {Diener}, {Distefano}, {Dolding}, {Drazinos}, {Dur{\'a}n},
  {Edvardsson}, {Enke}, {Eriksson}, {Esquej}, {Eynard Bontemps}, {Fabre},
  {Fabrizio}, {Faigler}, {Falc{\~a}o}, {Farr{\`a}s Casas}, {Federici},
  {Fedorets}, {Fernique}, {Figueras}, {Filippi}, {Findeisen}, {Fonti},
  {Fraile}, {Fraser}, {Fr{\'e}zouls}, {Gai}, {Galleti}, {Garabato},
  {Garc{\'\i}a-Sedano}, {Garofalo}, {Garralda}, {Gavel}, {Gavras}, {Gerssen},
  {Geyer}, {Giacobbe}, {Gilmore}, {Girona}, {Giuffrida}, {Glass}, {Gomes},
  {Granvik}, {Gueguen}, {Guerrier}, {Guiraud}, {Guti{\'e}rrez-S{\'a}nchez},
  {Haigron}, {Hatzidimitriou}, {Hauser}, {Haywood}, {Heiter}, {Helmi}, {Heu},
  {Hilger}, {Hobbs}, {Hofmann}, {Holland}, {Huckle}, {Hypki}, {Icardi},
  {Jan{\ss}en}, {Jevardat de Fombelle}, {Jonker}, {Juh{\'a}sz}, {Julbe},
  {Karampelas}, {Kewley}, {Klar}, {Kochoska}, {Kohley}, {Kolenberg},
  {Kontizas}, {Kontizas}, {Koposov}, {Kordopatis}, {Kostrzewa-Rutkowska},
  {Koubsky}, {Lambert}, {Lanza}, {Lasne}, {Lavigne}, {Le Fustec}, {Le
  Poncin-Lafitte}, {Lebreton}, {Leccia}, {Leclerc}, {Lecoeur-Taibi},
  {Lenhardt}, {Leroux}, {Liao}, {Licata}, {Lindstr{\o}m}, {Lister}, {Livanou},
  {Lobel}, {L{\'o}pez}, {Managau}, {Mann}, {Mantelet}, {Marchal}, {Marchant},
  {Marconi}, {Marinoni}, {Marschalk{\'o}}, {Marshall}, {Martino}, {Marton},
  {Mary}, {Massari}, {Matijevi{\v{c}}}, {Mazeh}, {McMillan}, {Messina},
  {Michalik}, {Millar}, {Molina}, {Molinaro}, {Moln{\'a}r}, {Montegriffo},
  {Mor}, {Morbidelli}, {Morel}, {Morris}, {Mulone}, {Muraveva}, {Musella},
  {Nelemans}, {Nicastro}, {Noval}, {O'Mullane}, {Ord{\'e}novic},
  {Ord{\'o}{\~n}ez-Blanco}, {Osborne}, {Pagani}, {Pagano}, {Pailler},
  {Palacin}, {Palaversa}, {Panahi}, {Pawlak}, {Piersimoni}, {Pineau}, {Plachy},
  {Plum}, {Poggio}, {Poujoulet}, {Pr{\v{s}}a}, {Pulone}, {Racero}, {Ragaini},
  {Rambaux}, {Ramos-Lerate}, {Regibo}, {Reyl{\'e}}, {Riclet}, {Ripepi}, {Riva},
  {Rivard}, {Rixon}, {Roegiers}, {Roelens}, {Romero-G{\'o}mez}, {Rowell},
  {Royer}, {Ruiz-Dern}, {Sadowski}, {Sagrist{\`a} Sell{\'e}s}, {Sahlmann},
  {Salgado}, {Salguero}, {Sanna}, {Santana-Ros}, {Sarasso}, {Savietto},
  {Schultheis}, {Sciacca}, {Segol}, {Segovia}, {S{\'e}gransan}, {Shih},
  {Siltala}, {Silva}, {Smart}, {Smith}, {Solano}, {Solitro}, {Sordo}, {Soria
  Nieto}, {Souchay}, {Spagna}, {Spoto}, {Stampa}, {Steele},
  {Steidelm{\"u}ller}, {Stephenson}, {Stoev}, {Suess}, {Surdej}, {Szabados},
  {Szegedi-Elek}, {Tapiador}, {Taris}, {Tauran}, {Taylor}, {Teixeira},
  {Terrett}, {Teyssand ier}, {Thuillot}, {Titarenko}, {Torra Clotet}, {Turon},
  {Ulla}, {Utrilla}, {Uzzi}, {Vaillant}, {Valentini}, {Valette}, {van Elteren},
  {Van Hemelryck}, {van Leeuwen}, {Vaschetto}, {Vecchiato}, {Veljanoski},
  {Viala}, {Vicente}, {Vogt}, {von Essen}, {Voss}, {Votruba}, {Voutsinas},
  {Walmsley}, {Weiler}, {Wertz}, {Wevers}, {Wyrzykowski}, {Yoldas},
  {{\v{Z}}erjal}, {Ziaeepour}, {Zorec}, {Zschocke}, {Zucker}, {Zurbach}, \&
  {Zwitter}}]{Gaia2018}
{Gaia Collaboration}, {Brown}, A.~G.~A., {Vallenari}, A., {et~al.} 2018, \aap,
  616, A1, \dodoi{10.1051/0004-6361/201833051}

\bibitem[{{Gandhi} \& {Madhusudhan}(2019)}]{gandhi2019}
{Gandhi}, S., \& {Madhusudhan}, N. 2019, \mnras, 485, 5817,
  \dodoi{10.1093/mnras/stz751}

\bibitem[{{Gazak} {et~al.}(2012){Gazak}, {Johnson}, {Tonry}, {Dragomir},
  {Eastman}, {Mann}, \& {Agol}}]{2012AdAst2012E..30G}
{Gazak}, J.~Z., {Johnson}, J.~A., {Tonry}, J., {et~al.} 2012, Advances in
  Astronomy, 2012, 697967, \dodoi{10.1155/2012/697967}

\bibitem[{{Gibson}(2014)}]{gibson14}
{Gibson}, N.~P. 2014, \mnras, 445, 3401, \dodoi{10.1093/mnras/stu1975}

\bibitem[{{Gibson} {et~al.}(2013{\natexlab{a}}){Gibson}, {Aigrain}, {Barstow},
  {Evans}, {Fletcher}, \& {Irwin}}]{gibson13a}
{Gibson}, N.~P., {Aigrain}, S., {Barstow}, J.~K., {et~al.} 2013{\natexlab{a}},
  \mnras, 428, 3680, \dodoi{10.1093/mnras/sts307}

\bibitem[{{Gibson} {et~al.}(2013{\natexlab{b}}){Gibson}, {Aigrain}, {Barstow},
  {Evans}, {Fletcher}, \& {Irwin}}]{gibson13b}
---. 2013{\natexlab{b}}, \mnras, 436, 2974, \dodoi{10.1093/mnras/stt1783}

\bibitem[{{Gibson} {et~al.}(2012){Gibson}, {Aigrain}, {Roberts}, {Evans},
  {Osborne}, \& {Pont}}]{gibson12a}
{Gibson}, N.~P., {Aigrain}, S., {Roberts}, S., {et~al.} 2012, \mnras, 419,
  2683, \dodoi{10.1111/j.1365-2966.2011.19915.x}

\bibitem[{{Gibson} {et~al.}(2017){Gibson}, {Nikolov}, {Sing}, {Barstow},
  {Evans}, {Kataria}, \& {Wilson}}]{gibson17}
{Gibson}, N.~P., {Nikolov}, N., {Sing}, D.~K., {et~al.} 2017, \mnras, 467,
  4591, \dodoi{10.1093/mnras/stx353}

\bibitem[{{Gillon} {et~al.}(2009){Gillon}, {Anderson}, {Triaud}, {Hellier},
  {Maxted}, {Pollaco}, {Queloz}, {Smalley}, {West}, {Wilson}, {Bentley},
  {Collier Cameron}, {Enoch}, {Hebb}, {Horne}, {Irwin}, {Joshi}, {Lister},
  {Mayor}, {Pepe}, {Parley}, {Segransan}, {Udry}, \& {Wheatley}}]{Gillon09}
{Gillon}, M., {Anderson}, D.~R., {Triaud}, A.~H.~M.~J., {et~al.} 2009, \aap,
  501, 785, \dodoi{10.1051/0004-6361/200911749}

\bibitem[{{Huitson} {et~al.}(2017){Huitson}, {D{\'e}sert}, {Bean}, {Fortney},
  {Stevenson}, \& {Bergmann}}]{Huitson17}
{Huitson}, C.~M., {D{\'e}sert}, J.~M., {Bean}, J.~L., {et~al.} 2017, \aj, 154,
  95, \dodoi{10.3847/1538-3881/aa7f72}

\bibitem[{{Huitson} {et~al.}(2013){Huitson}, {Sing}, {Pont}, {Fortney},
  {Burrows}, {Wilson}, {Ballester}, {Nikolov}, {Gibson}, {Deming}, {Aigrain},
  {Evans}, {Henry}, {Lecavelier des Etangs}, {Showman}, {Vidal-Madjar}, \&
  {Zahnle}}]{huitson13}
{Huitson}, C.~M., {Sing}, D.~K., {Pont}, F., {et~al.} 2013, \mnras, 434, 3252,
  \dodoi{10.1093/mnras/stt1243}

\bibitem[{{Husser} {et~al.}(2013){Husser}, {Wende-von Berg}, {Dreizler},
  {Homeier}, {Reiners}, {Barman}, \& {Hauschildt}}]{Husser2013}
{Husser}, T.~O., {Wende-von Berg}, S., {Dreizler}, S., {et~al.} 2013, \aap,
  553, A6, \dodoi{10.1051/0004-6361/201219058}

\bibitem[{{Jayasinghe} {et~al.}(2019){Jayasinghe}, {Stanek}, {Kochanek},
  {Shappee}, {Holoien}, {Thompson}, {Prieto}, {Dong}, {Pawlak}, {Pejcha},
  {Shields}, {Pojmanski}, {Otero}, {Hurst}, {Britt}, \&
  {Will}}]{Jayasinghe2019}
{Jayasinghe}, T., {Stanek}, K.~Z., {Kochanek}, C.~S., {et~al.} 2019, \mnras,
  485, 961, \dodoi{10.1093/mnras/stz444}

\bibitem[{{Kirk} {et~al.}(2018){Kirk}, {Wheatley}, {Louden}, {Skillen}, {King},
  {McCormac}, \& {Irwin}}]{Kirk18}
{Kirk}, J., {Wheatley}, P.~J., {Louden}, T., {et~al.} 2018, \mnras, 474, 876,
  \dodoi{10.1093/mnras/stx2826}

\bibitem[{{Kirkby}(2007)}]{2007SGeo...28..333K}
{Kirkby}, J. 2007, Surveys in Geophysics, 28, 333,
  \dodoi{10.1007/s10712-008-9030-6}

\bibitem[{{Kirkby} {et~al.}(2011){Kirkby}, {Curtius}, {Almeida}, {Dunne},
  {Duplissy}, {Ehrhart}, {Franchin}, {Gagn{\'e}}, {Ickes}, {K{\"u}rten},
  {Kupc}, {Metzger}, {Riccobono}, {Rondo}, {Schobesberger}, {Tsagkogeorgas},
  {Wimmer}, {Amorim}, {Bianchi}, {Breitenlechner}, {David}, {Dommen},
  {Downard}, {Ehn}, {Flagan}, {Haider}, {Hansel}, {Hauser}, {Jud}, {Junninen},
  {Kreissl}, {Kvashin}, {Laaksonen}, {Lehtipalo}, {Lima}, {Lovejoy},
  {Makhmutov}, {Mathot}, {Mikkil{\"a}}, {Minginette}, {Mogo}, {Nieminen},
  {Onnela}, {Pereira}, {Pet{\"a}j{\"a}}, {Schnitzhofer}, {Seinfeld},
  {Sipil{\"a}}, {Stozhkov}, {Stratmann}, {Tom{\'e}}, {Vanhanen}, {Viisanen},
  {Vrtala}, {Wagner}, {Walther}, {Weingartner}, {Wex}, {Winkler}, {Carslaw},
  {Worsnop}, {Baltensperger}, \& {Kulmala}}]{2011Natur.476..429K}
{Kirkby}, J., {Curtius}, J., {Almeida}, J., {et~al.} 2011, \nat, 476, 429,
  \dodoi{10.1038/nature10343}

\bibitem[{{Kreidberg}(2015)}]{kreidberg2015}
{Kreidberg}, L. 2015, \pasp, 127, 1161, \dodoi{10.1086/683602}

\bibitem[{{Lightkurve Collaboration} {et~al.}(2018){Lightkurve Collaboration},
  {Cardoso}, {Hedges}, {Gully-Santiago}, {Saunders}, {Cody}, {Barclay}, {Hall},
  {Sagear}, {Turtelboom}, {Zhang}, {Tzanidakis}, {Mighell}, {Coughlin}, {Bell},
  {Berta-Thompson}, {Williams}, {Dotson}, \& {Barentsen}}]{2018ascl.soft12013L}
{Lightkurve Collaboration}, {Cardoso}, J. V. d. M.~a., {Hedges}, C., {et~al.}
  2018, {Lightkurve: Kepler and TESS time series analysis in Python}.
\newblock \doeprint{1812.013}

\bibitem[{{Lines} {et~al.}(2018){Lines}, {Mayne}, {Boutle}, {Manners}, {Lee},
  {Helling}, {Drummond}, {Amundsen}, {Goyal}, {Acreman}, {Tremblin}, \&
  {Kerslake}}]{2018A&A...615A..97L}
{Lines}, S., {Mayne}, N.~J., {Boutle}, I.~A., {et~al.} 2018, \aap, 615, A97,
  \dodoi{10.1051/0004-6361/201732278}

\bibitem[{{Lodders}(1999)}]{lodders99}
{Lodders}, K. 1999, \apj, 519, 793, \dodoi{10.1086/307387}

\bibitem[{{Lodders}(2002)}]{lodders02a}
---. 2002, \apj, 577, 974, \dodoi{10.1086/342241}

\bibitem[{{Madhusudhan} \& {Seager}(2009)}]{Madhusudhan2009}
{Madhusudhan}, N., \& {Seager}, S. 2009, \apj, 707, 24,
  \dodoi{10.1088/0004-637X/707/1/24}

\bibitem[{{Magic} {et~al.}(2015){Magic}, {Chiavassa}, {Collet}, \&
  {Asplund}}]{magic2015}
{Magic}, Z., {Chiavassa}, A., {Collet}, R., \& {Asplund}, M. 2015, \aap, 573,
  A90, \dodoi{10.1051/0004-6361/201423804}

\bibitem[{{Mandel} \& {Agol}(2002)}]{mandel02}
{Mandel}, K., \& {Agol}, E. 2002, \apjl, 580, L171, \dodoi{10.1086/345520}

\bibitem[{{Markwardt}(2009)}]{markwardt09}
{Markwardt}, C.~B. 2009, in Astronomical Society of the Pacific Conference
  Series, Vol. 411, Astronomical Data Analysis Software and Systems XVIII, ed.
  D.~A. {Bohlender}, D.~{Durand}, \& P.~{Dowler}, 251.
\newblock \doarXiv{0902.2850}

\bibitem[{{McCullough} {et~al.}(2014){McCullough}, {Crouzet}, {Deming}, \&
  {Madhusudhan}}]{McCullough14}
{McCullough}, P.~R., {Crouzet}, N., {Deming}, D., \& {Madhusudhan}, N. 2014,
  \apj, 791, 55, \dodoi{10.1088/0004-637X/791/1/55}

\bibitem[{{Nikolov} {et~al.}(2016){Nikolov}, {Sing}, {Gibson}, {Fortney},
  {Evans}, {Barstow}, {Kataria}, \& {Wilson}}]{nikolov16}
{Nikolov}, N., {Sing}, D.~K., {Gibson}, N.~P., {et~al.} 2016, \apj, 832, 191,
  \dodoi{10.3847/0004-637X/832/2/191}

\bibitem[{{Nikolov} {et~al.}(2018){Nikolov}, {Sing}, {Fortney}, {Goyal},
  {Drummond}, {Evans}, {Gibson}, {De Mooij}, {Rustamkulov}, {Wakeford},
  {Smalley}, {Burgasser}, {Hellier}, {Helling}, {Mayne}, {Madhusudhan},
  {Kataria}, {Baines}, {Carter}, {Ballester}, {Barstow}, {McCleery}, \&
  {Spake}}]{nikolov18b}
{Nikolov}, N., {Sing}, D.~K., {Fortney}, J.~J., {et~al.} 2018, \nat, 557, 526,
  \dodoi{10.1038/s41586-018-0101-7}

\bibitem[{{Noyes} {et~al.}(1984){Noyes}, {Hartmann}, {Baliunas}, {Duncan}, \&
  {Vaughan}}]{1984ApJ...279..763N}
{Noyes}, R.~W., {Hartmann}, L.~W., {Baliunas}, S.~L., {Duncan}, D.~K., \&
  {Vaughan}, A.~H. 1984, \apj, 279, 763, \dodoi{10.1086/161945}

\bibitem[{{Palle} {et~al.}(2017){Palle}, {Chen}, {Prieto-Arranz}, {Nowak},
  {Murgas}, {Nortmann}, {Pollacco}, {Lam}, {Montanes-Rodriguez}, {Parviainen},
  \& {Casasayas-Barris}}]{Palle17}
{Palle}, E., {Chen}, G., {Prieto-Arranz}, J., {et~al.} 2017, \aap, 602, L15,
  \dodoi{10.1051/0004-6361/201731018}

\bibitem[{{Pinhas} {et~al.}(2018){Pinhas}, {Rackham}, {Madhusudhan}, \&
  {Apai}}]{Pinhas2018}
{Pinhas}, A., {Rackham}, B.~V., {Madhusudhan}, N., \& {Apai}, D. 2018, \mnras,
  480, 5314, \dodoi{10.1093/mnras/sty2209}

\bibitem[{{Pollacco} {et~al.}(2006){Pollacco}, {Skillen}, {Collier Cameron},
  {Christian}, {Hellier}, {Irwin}, {Lister}, {Street}, {West}, {Anderson},
  {Clarkson}, {Deeg}, {Enoch}, {Evans}, {Fitzsimmons}, {Haswell}, {Hodgkin},
  {Horne}, {Kane}, {Keenan}, {Maxted}, {Norton}, {Osborne}, {Parley}, {Ryans},
  {Smalley}, {Wheatley}, \& {Wilson}}]{2006PASP..118.1407P}
{Pollacco}, D.~L., {Skillen}, I., {Collier Cameron}, A., {et~al.} 2006, \pasp,
  118, 1407, \dodoi{10.1086/508556}

\bibitem[{{Pont} {et~al.}(2013){Pont}, {Sing}, {Gibson}, {Aigrain}, {Henry}, \&
  {Husnoo}}]{pont13}
{Pont}, F., {Sing}, D.~K., {Gibson}, N.~P., {et~al.} 2013, \mnras, 432, 2917,
  \dodoi{10.1093/mnras/stt651}

\bibitem[{{Pont} {et~al.}(2006){Pont}, {Zucker}, \& {Queloz}}]{pont06}
{Pont}, F., {Zucker}, S., \& {Queloz}, D. 2006, \mnras, 373, 231,
  \dodoi{10.1111/j.1365-2966.2006.11012.x}

\bibitem[{{Rackham} {et~al.}(2017){Rackham}, {Espinoza}, {Apai},
  {L{\'o}pez-Morales}, {Jord{\'a}n}, {Osip}, {Lewis}, {Rodler}, {Fraine},
  {Morley}, \& {Fortney}}]{Rackham2017}
{Rackham}, B., {Espinoza}, N., {Apai}, D., {et~al.} 2017, \apj, 834, 151,
  \dodoi{10.3847/1538-4357/aa4f6c}

\bibitem[{{Ricker} {et~al.}(2014){Ricker}, {Winn}, {Vanderspek}, {Latham},
  {Bakos}, {Bean}, {Berta-Thompson}, {Brown}, {Buchhave}, {Butler}, {Butler},
  {Chaplin}, {Charbonneau}, {Christensen-Dalsgaard}, {Clampin}, {Deming},
  {Doty}, {De Lee}, {Dressing}, {Dunham}, {Endl}, {Fressin}, {Ge}, {Henning},
  {Holman}, {Howard}, {Ida}, {Jenkins}, {Jernigan}, {Johnson}, {Kaltenegger},
  {Kawai}, {Kjeldsen}, {Laughlin}, {Levine}, {Lin}, {Lissauer}, {MacQueen},
  {Marcy}, {McCullough}, {Morton}, {Narita}, {Paegert}, {Palle}, {Pepe},
  {Pepper}, {Quirrenbach}, {Rinehart}, {Sasselov}, {Sato}, {Seager},
  {Sozzetti}, {Stassun}, {Sullivan}, {Szentgyorgyi}, {Torres}, {Udry}, \&
  {Villasenor}}]{2014SPIE.9143E..20R}
{Ricker}, G.~R., {Winn}, J.~N., {Vanderspek}, R., {et~al.} 2014, in Society of
  Photo-Optical Instrumentation Engineers (SPIE) Conference Series, Vol. 9143,
  Space Telescopes and Instrumentation 2014: Optical, Infrared, and Millimeter
  Wave, ed. J.~{Oschmann}, Jacobus~M., M.~{Clampin}, G.~G. {Fazio}, \& H.~A.
  {MacEwen}, 914320, \dodoi{10.1117/12.2063489}

\bibitem[{{Roeser} {et~al.}(2010){Roeser}, {Demleitner}, \&
  {Schilbach}}]{roeser2010}
{Roeser}, S., {Demleitner}, M., \& {Schilbach}, E. 2010, \aj, 139, 2440,
  \dodoi{10.1088/0004-6256/139/6/2440}

\bibitem[{Rothman {et~al.}(2010)Rothman, Gordon, Barber, Dothe, Gamache,
  Goldman, Perevalov, Tashkun, \& Tennyson}]{Rothman2010}
Rothman, L., Gordon, I., Barber, R., {et~al.} 2010, \jqsrt, 111, 2139 ,
  \dodoi{https://doi.org/10.1016/j.jqsrt.2010.05.001}

\bibitem[{Schwarz(1978)}]{Schwarz78}
Schwarz, G. 1978, Annals of Statistics, 6, 461

\bibitem[{{Seager} \& {Sasselov}(2000)}]{seager00}
{Seager}, S., \& {Sasselov}, D.~D. 2000, \apj, 537, 916, \dodoi{10.1086/309088}

\bibitem[{{Shappee} {et~al.}(2014){Shappee}, {Prieto}, {Grupe}, {Kochanek},
  {Stanek}, {De Rosa}, {Mathur}, {Zu}, {Peterson}, {Pogge}, {Komossa}, {Im},
  {Jencson}, {Holoien}, {Basu}, {Beacom}, {Szczygie{\l}}, {Brimacombe},
  {Adams}, {Campillay}, {Choi}, {Contreras}, {Dietrich}, {Dubberley},
  {Elphick}, {Foale}, {Giustini}, {Gonzalez}, {Hawkins}, {Howell}, {Hsiao},
  {Koss}, {Leighly}, {Morrell}, {Mudd}, {Mullins}, {Nugent}, {Parrent},
  {Phillips}, {Pojmanski}, {Rosing}, {Ross}, {Sand}, {Terndrup}, {Valenti},
  {Walker}, \& {Yoon}}]{Shappee2014}
{Shappee}, B.~J., {Prieto}, J.~L., {Grupe}, D., {et~al.} 2014, \apj, 788, 48,
  \dodoi{10.1088/0004-637X/788/1/48}

\bibitem[{{Sing} {et~al.}(2016){Sing}, {Fortney}, {Nikolov}, {Wakeford},
  {Kataria}, {Evans}, {Aigrain}, {Ballester}, {Burrows}, {Deming},
  {D{\'e}sert}, {Gibson}, {Henry}, {Huitson}, {Knutson}, {Lecavelier Des
  Etangs}, {Pont}, {Showman}, {Vidal-Madjar}, {Williamson}, \&
  {Wilson}}]{sing16}
{Sing}, D.~K., {Fortney}, J.~J., {Nikolov}, N., {et~al.} 2016, \nat, 529, 59,
  \dodoi{10.1038/nature16068}

\bibitem[{{Sing} {et~al.}(2019){Sing}, {Lavvas}, {Ballester}, {Lecavelier des
  Etangs}, {Marley}, {Nikolov}, {Ben-Jaffel}, {Bourrier}, {Buchhave}, {Deming},
  {Ehrenreich}, {Mikal-Evans}, {Kataria}, {Lewis}, {L{\'o}pez-Morales},
  {Garc{\'\i}a Mu{\~n}oz}, {Henry}, {Sanz-Forcada}, {Spake}, {Wakeford}, \&
  {PanCET Collaboration}}]{Sing19}
{Sing}, D.~K., {Lavvas}, P., {Ballester}, G.~E., {et~al.} 2019, \aj, 158, 91,
  \dodoi{10.3847/1538-3881/ab2986}

\bibitem[{{Su{\'a}rez Mascare{\~n}o} {et~al.}(2015){Su{\'a}rez Mascare{\~n}o},
  {Rebolo}, {Gonz{\'a}lez Hern{\'a}ndez}, \& {Esposito}}]{2015MNRAS.452.2745S}
{Su{\'a}rez Mascare{\~n}o}, A., {Rebolo}, R., {Gonz{\'a}lez Hern{\'a}ndez},
  J.~I., \& {Esposito}, M. 2015, \mnras, 452, 2745,
  \dodoi{10.1093/mnras/stv1441}

\bibitem[{{Svensmark}(1998)}]{1998PhRvL..81.5027S}
{Svensmark}, H. 1998, \prl, 81, 5027, \dodoi{10.1103/PhysRevLett.81.5027}

\bibitem[{{Svensmark} {et~al.}(2009){Svensmark}, {Bondo}, \&
  {Svensmark}}]{2009GeoRL..3615101S}
{Svensmark}, H., {Bondo}, T., \& {Svensmark}, J. 2009, \grl, 36, L15101,
  \dodoi{10.1029/2009GL038429}

\bibitem[{{Tody}(1986)}]{1986SPIE..627..733T}
{Tody}, D. 1986, in Society of Photo-Optical Instrumentation Engineers (SPIE)
  Conference Series, Vol. 627, Instrumentation in astronomy VI, ed. D.~L.
  {Crawford}, 733, \dodoi{10.1117/12.968154}

\bibitem[{{Tody}(1993)}]{1993ASPC...52..173T}
{Tody}, D. 1993, in Astronomical Society of the Pacific Conference Series,
  Vol.~52, Astronomical Data Analysis Software and Systems II, ed. R.~J.
  {Hanisch}, R.~J.~V. {Brissenden}, \& J.~{Barnes}, 173

\bibitem[{{Vaughan} {et~al.}(1978){Vaughan}, {Preston}, \&
  {Wilson}}]{1978PASP...90..267V}
{Vaughan}, A.~H., {Preston}, G.~W., \& {Wilson}, O.~C. 1978, \pasp, 90, 267,
  \dodoi{10.1086/130324}

\bibitem[{{Wakeford} \& {Sing}(2015)}]{2015A&A...573A.122W}
{Wakeford}, H.~R., \& {Sing}, D.~K. 2015, \aap, 573, A122,
  \dodoi{10.1051/0004-6361/201424207}

\bibitem[{{Welbanks} {et~al.}(2019){Welbanks}, {Madhusudhan}, {Allard},
  {Hubeny}, {Spiegelman}, \& {Leininger}}]{Welbanks2019b}
{Welbanks}, L., {Madhusudhan}, N., {Allard}, N.~F., {et~al.} 2019, ApJLett,
  887, L20, \dodoi{10.3847/2041-8213/ab5a89}

\bibitem[{{Wilson} {et~al.}(2020){Wilson}, {Gibson}, \& {Nikolov}}]{wilson20}
{Wilson}, J., {Gibson}, N.~P., \& {Nikolov}, N. 2020, \mnras, 999, 9999

\bibitem[{{Wyttenbach} {et~al.}(2017){Wyttenbach}, {Lovis}, {Ehrenreich},
  {Bourrier}, {Pino}, {Allart}, {Astudillo-Defru}, {Cegla}, {Heng}, {Lavie},
  {Melo}, {Murgas}, {Santerne}, {S{\'e}gransan}, {Udry}, \&
  {Pepe}}]{Wyttenbach17}
{Wyttenbach}, A., {Lovis}, C., {Ehrenreich}, D., {et~al.} 2017, \aap, 602, A36,
  \dodoi{10.1051/0004-6361/201630063}

\end{thebibliography}
\bibliographystyle{aasjournal}

\appendix

\section{Spectroscopic light curves}\label{app_spec_lcs}

\begin{figure*}[h]
\begin{centering}
\includegraphics[width=0.82\textwidth]{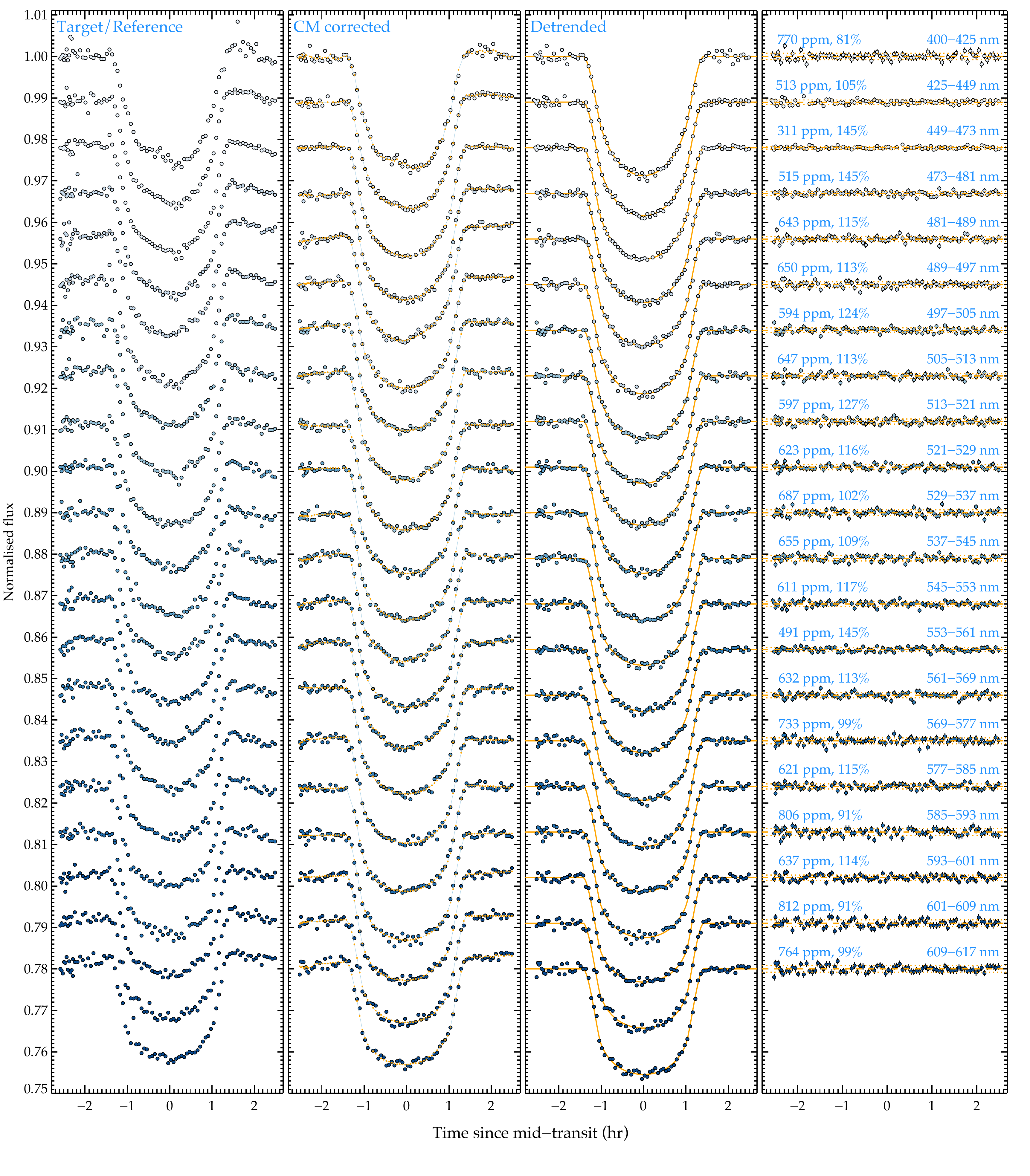}
\caption{Spectrophotometric light curves from grism 600B offset by a constant amount for clarity. The first panel shows the raw target-to-reference flux. The second panel shows the common-mode (CM)-corrected light curves and the transit and systematics models, with the highest statistical weight. The third panel shows systematics corrected light curves and the transit model with the highest statistical weight. The fourth panel shows residuals with 1$\sigma$ error bars. The dashed lines indicate the median residual level, with dotted lines indicating the dispersion and the percentage of the theoretical photon noise limit reached (blue) \label{fig:lcs_b}}
\end{centering}
\end{figure*}

\begin{figure*}[h]
\begin{centering}
\includegraphics[width=0.82\textwidth]{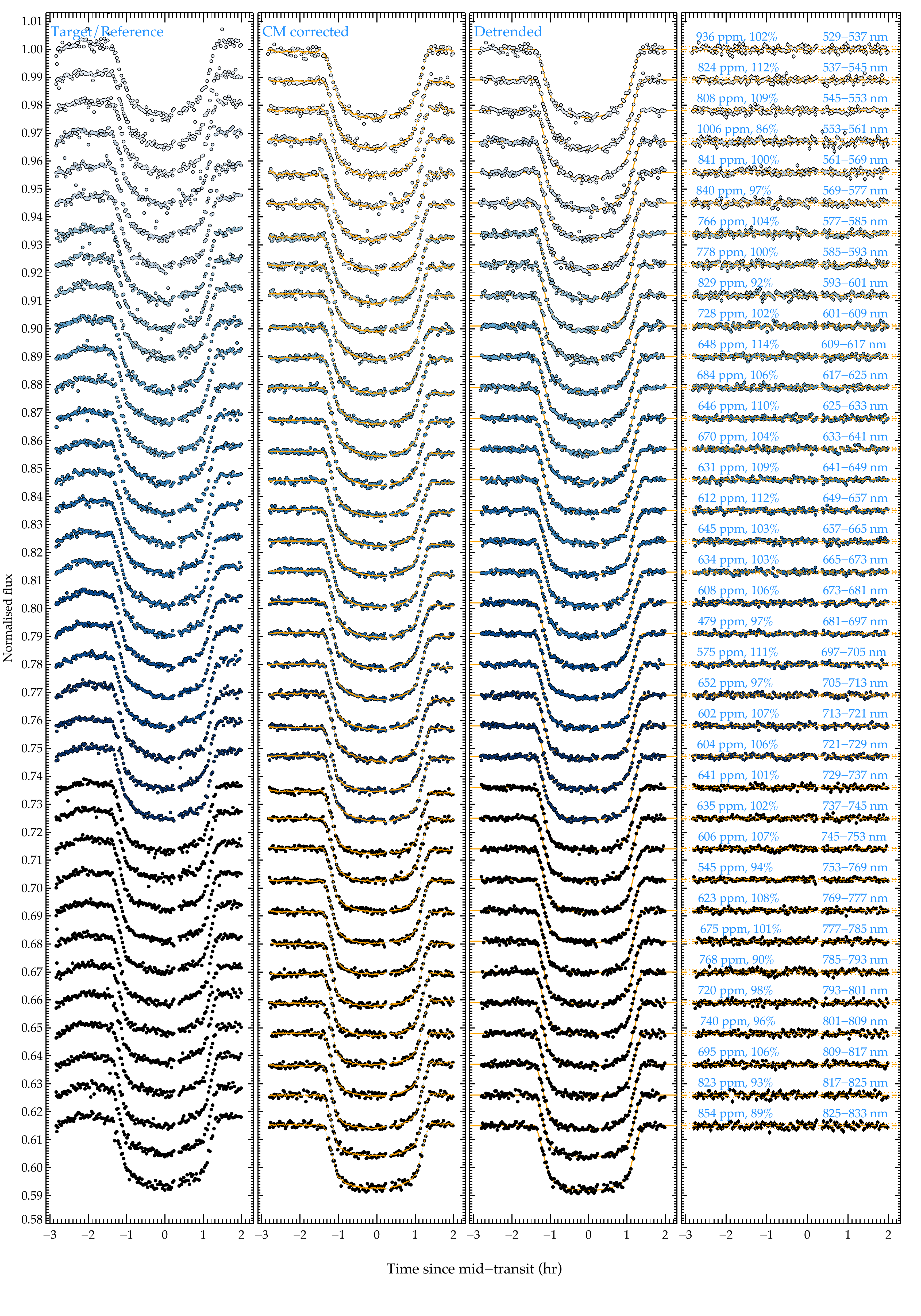}
\caption{As for Figure\,\ref{fig:lcs_b} but for grism 600RI.  \label{fig:lcs_ri}}
\end{centering}
\end{figure*}

\section{Stellar Variability}\label{app_variab}

\begin{figure*}[h]
\includegraphics[width=1\textwidth]{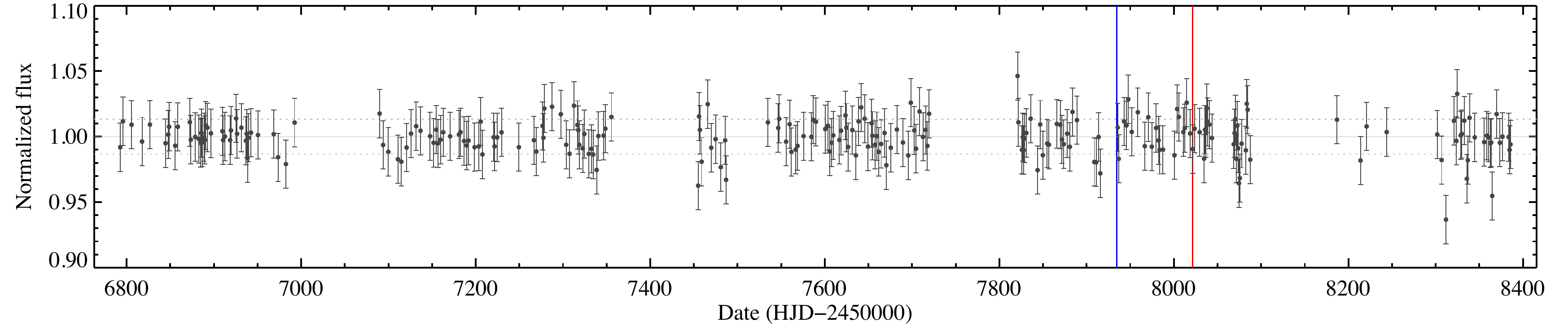}
\caption{ASAS-SN light curve of WASP-110. The error bars indicate $1\sigma$ uncertainties. The vertical lines indicate the FORS2 GRIS600B (blue) and GRIS600RI (red) transit epocs, respectively. The horizontal lines indicate the mean and $1\sigma$ levels.  \label{fig:varLC1}}
\end{figure*}

\section{Retrieval results}\label{app_retr}

\begin{figure*}[h]
    \centering
    \includegraphics[width=\columnwidth]{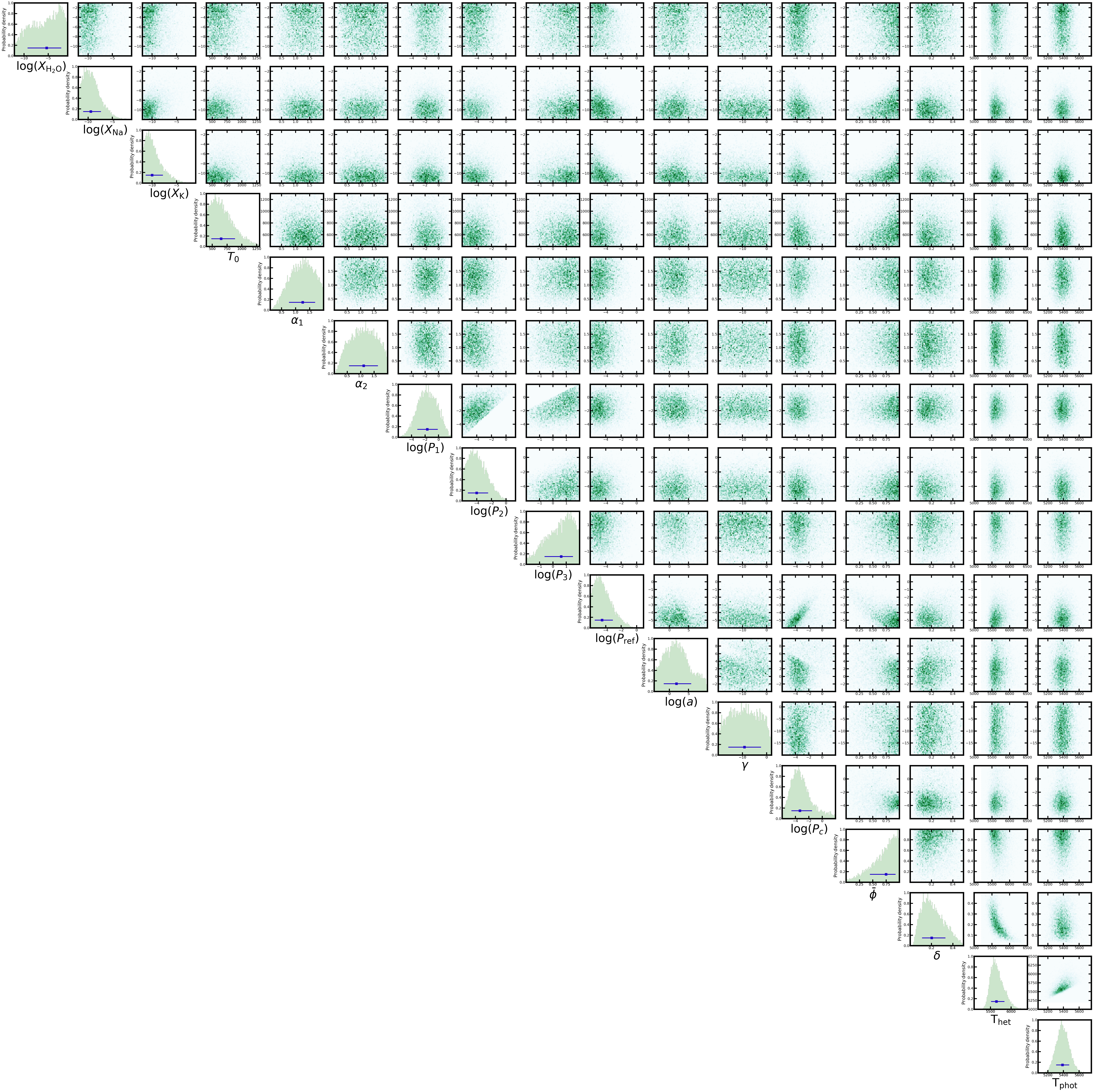}
    \caption{Posterior distribution obtained from our AURA retrieval, consisting of 3 chemical abundances, 6 pressure-temperature profile parameters, the reference pressure at the planetary radius, 4 clouds and hazes parameters and 3 stellar heterogeneity parameters.}
    \label{fig:AURA_full_posterior}
\end{figure*}



\end{document}